\documentclass[preprint2,english]{aastex}
\usepackage{graphicx}
\usepackage[T1]{fontenc}
\usepackage[latin9]{inputenc}
\usepackage{array}
\usepackage{float}
\usepackage{graphicx}
\usepackage{amssymb}

\makeatletter

\providecommand{\tabularnewline}{\\}

\@ifundefined{showcaptionsetup}{}{%
 \PassOptionsToPackage{caption=false}{subfig}}
\usepackage{subfig}
\makeatother

\usepackage{babel}

\begin{document}
\title{Measuring the pulse of GRB 090618: A Simultaneous Spectral and Timing
Analysis of the Prompt Emission}

\author{Rupal Basak$\rm^{1}$ and A. R. Rao$\rm^{2}$}
\affil{Tata Institute of Fundamental Research, Mumbai - 400005, India.}

\altaffiltext{1}{$rupalb@tifr.res.in$}
\altaffiltext{2}{$arrao@tifr.res.in$}

\clearpage

\makeatletter

%\email{rupalb@tifr.res.in}
\begin{abstract}
We develop a new method for simultaneous timing and spectral studies of Gamma Ray 
Burst (GRB) prompt emission and apply it to make a
pulse-wise description of the prompt emission of 
GRB~090618, the brightest GRB detected in the Fermi era. 
%make a simultaneous spectral measurement using the \emph{Swift}/BAT and
%\emph{Fermi}/GBM data of 
We exploit the large area (and sensitivity)
of \emph{Swift}/BAT and the wide band width of \emph{Fermi}/GBM 
to derive the parameters for a complete spectral and timing description of the
individual pulses of this GRB, based on the various empirical relations suggested in
the literature. We demonstrate that this empirical model correctly
describes the other observed properties of the burst like the variation of the
lag with energy  and the 
pulse width with energy. 
The measurements also show an indication of an \emph{increase} in pulse width
as a function of energy at low energies for some of the pulses, which is naturally
explained as an off-shoot of some particular combination of the 
model parameters. 
%We develop a method
% In particular,  the model parameters
%allow for an increase or des
%
%significantly improve the spectral parameters. The data also
%allow the examination of individual pulses of this GRB at finer energy
%bins and we detect a peculiar deviation in the dependence of pulse
%width with energy: instead of the canonical power-law decrease of
%width with energy, the measurements also show an \emph{increase} in width
%as a function of energy at low energies for some of the pulses. We develop a method
%to make a simultaneous fit to the energy spectrum and pulse profile
%of individual pulses and show that the spectral and timing characteristics
%can be explained in terms of a few ($\sim$ 2) model parameters. The dependence
We argue that these model parameters, particularly the peak energy at the
beginning of the pulse, are the natural choices to be used for 
correlation with luminosity. 
%of these parameters with luminosity shows that the individual pulses
%of GRBs can be used as distance indicators. 
The implications of these
results for the use of GRBs as standard candles are briefly described.
\end{abstract}


\section{INTRODUCTION}

The phenomenon of Gamma Ray Burst (GRB) has raised many unresolved issues.
For example, the nature of the compact object in the progenitor responsible for the
jet emission is not yet understood. 
The widely accepted scenario for the GRB emission, supported by strong observational evidences
 (van Paradijs et al. 2000),
is that of a catastrophic energy release from a highly massive, rapidly rotating,
low metallicity star towards the end of its life. This catastrophic process for 
the long GRB class (T$_{90}$ $\gtrsim$ 2 s, where T$_{90}$ is the time
taken by the burst to accumulate between 5\% to 95\% of its prompt 
emission -- see Kouveliotou et al. 1993; Norris et al. 2005) is attributed to the collapsar or
hypernova scenario (Woosley 1993; Paczynski 1998; Fryer et
al. 1999; M{\'e}sz{\'a}ros 2006) while the short GRBs (T$_{90}$ $\lesssim$
2 s) are thought of as the outcome of merging neutron star  binaries
 or neutron star-black hole  binaries (Paczynski
1986; Meszaros 1992;1997; Rosswog et al. 2003a; b). Though the collapsar model
supports the formation of a stellar mass black hole, it is also
suggested that  the compact object could be a
magnetar (Metzger et al. 2007; Metzger 2010). 

The prompt emission of long GRBs shows a wide range of structures in their
light curve, though the time integrated energy spectra are adequately
fit by a four parameter Band model (Band et al. 1993). Attempts have
been made to fit the light curve of the prompt emission with
empirical models such as the Fast Rise Exponential Decay (FRED) model
(Kocevski et al. 2003) and the exponential model (Norris et al. 2005). These
models assume an underlying pulse structure of the prompt emission.
The curve fitting shows that the pulses are nearly simultaneous in
a wide range of energy bands (Norris et al. 2005) and have self similar
shapes (Nemiroff 2000). Hakkila et al. (2011) showed that the observed
properties of the short pulses of long GRBs correlate the same way as
typical pulses of short GRBs. All these properties of pulse emission
indicate that despite being diverse in their
many features, the prompt emission of GRB has a simple underlying
mechanism of pulse emission which can be characterized by empirical
models irrespective of the progenitor or the environment of its formation
(Hakkila et al. 2011).

The properties of the prompt emission of GRBs like $E_{peak}$ (the energy at which 
the emission peaks -- Amati et al. 2002), the lag between the light curves at
two energies (Norris et al. 2000),
 the average variability seen during the prompt emission (Fenimore \& Ramirez-Ruiz 2000),
 the rise time of the pulses (Schaefer 2002) are found to be
correlated to the luminosity of the bursts
and hence they are used as distance indicators (Schaefer 2007; 
Gehrels et al. 2009).  Most of these derived relations are empirical in nature
and the  analyses that are done to derive these
         parameters, however, are done separately in the time or the energy domain. 
         For example, to derive $E_{peak}$ the time-integrated spectrum is used, ignoring
         the time variability and the pulse characteristics. Similarly, to derive lags, 
the time profile is examined
   in a few broad energy bands. Attempts have been made to investigate the variations of
these derived parameters in the other domain. For example, the variation of $E_{peak}$
with time has been investigated by Kocevski \& Liang (2003). A complete empirical 
description of GRB prompt emission which describes the time as well as energy distribution,
however, is lacking. This is probably due to the method of spectral parameter determination
which is forward in nature: that is one assumes a possible spectral model, convolves with 
the detector response and verifies the validity of the model by comparing with the data
(Arnaud 1996). Hence it is difficult to incorporate time variability parameters in this
scheme. GRBs often have overlapping pulses which further complicates the extracting of
model parameters for any unified empirical description.

In this paper we develop a technique to give a complete pulse-wise
empirical description of the prompt emission of a GRB using the
time-integrated spectrum and the energy integrated light curve. 
%
%
%Pulse analysis can also provide a minimal number of parameters needed
%to describe various correlations in time resolved spectral study as
%opposed to the time division guided by intensity variation of the
%GRB as a whole (see for example, Ghirlanda et al. 2010). Moreover,
%these correlations and spectral parameters which are very important
%in characterizing the spectrum of the prompt emission of the burst,
%need to be measured very precisely and must not be influenced by instrumental
%systematics. One way to minimize the error in measurement of the spectral
%parameters is to do a joint analysis using two or more satellite data
%(see for example Sakamoto et al. 2011).
%
%With the dual aim of reducing the systematic errors in the spectral
%fitting and to make a detailed pulse-wise spectral and timing analysis,
We apply   this method to the \emph{Fermi}/GBM and the \emph{Swift}/BAT data of 
GRB 090618,
%. This GRB was simultaneously observed with \emph{Swift}/BAT
%(Schady et al. 2009a, 2009b, 2009c),\emph{ AGILE} (Longo et al. 2009),
%\emph{Fermi}/GBM (McBreen et al. 2009), \emph{Suzaku}/WAM (Kono et
%al.), \emph{Wind}/Konus-Wind, \emph{Coronas-Photon}/Konus-RF (Golenetskii
%et al. 2009), \emph{Coronas-Photon}/RT-2 (Rao et al. 2009). This is
%a long GRB (T$_{90}$$\sim$113s) with a fluence of $3398.1\pm62.00\times10^{-7}$
%erg/cm$^{2}$ (integrated over $\triangle t$=182.27s) 
which has the
highest fluence among all 438 GBM GRBs until March 2010 (Nava et al. 2011).
% (The
%next one being GRB~090902(462) with a fluence of $3223.60\pm28.46.00\times10^{-7}$erg/cm$^{2}$).
%Further, this GRB has a multi-peaked light curve which makes it suitable
%for investigation of various correlations in a single GRB.
%
We exploit the large area of \emph{Swift}/BAT to get accurate light curves and the
large band-width of \emph{Fermi}/GBM to get accurate energy spectra. 
%In this work we make a time resolved spectral analysis, similar to
%the work reported by Ghirlanda et al. (2010) and use the \emph{Fermi}/GBM data
%along with the \emph{Swift}/BAT data to show quantitatively that inclusion of more than
%one instrument lead to minimize the error in measured quantities despite
%the fact that different instrument has different systematic error. The light curves 
%of the GRB from data of both satellites are extracted
%for different energy bands and fitted with Norris' exponential
%model. The width and the asymmetry of each pulse are derived. 
%The measured pulse width shows an \emph{increase} from the lower energy 
%band to some of the higher energy bands for two pulses. The possible 
%explanations are given. The spectral analysis for each pulse is done with a Band function
%and the global values of the parameters are estimated for each pulse.
We use the Band description for the time integrated spectral model (Band 2003) and the exponential 
model (Norris et al. 2005) for the energy integrated timing description for the pulses.
We use the expression for the variation of $E_{peak}$ with time given by Kocevski \& Liang (2003)
and develop a method to derive the model parameters (see section 4.1.2). 
We demonstrate that these model parameters correctly describe the 
variation of pulse width with energy and the variation of the lag with energy.

In \S 2 we give a brief description of GRB~090618, and in \S 3 we give the data
analysis techniques including spectral analysis (\S 3.1), timing analysis (\S 3.2) and
a pulse-wise joint (\emph{Swift}/BAT and \emph{Fermi}/GBM) spectral analysis (\S 3.3). In \S 4
we describe our method in detail, which gives a complete three dimensional pulse-wise
description of the GRB. The model predicted timing results are compared with the
observations  in \S 4.2 and the implications for global correlations are touched upon
in \S 4.3. The results are discussed and the major conclusions are given in the
last section (\S 5). 
%Using these model predicted light curves we calculate some useful 
%timing parameters like delay, pulse width and observe the 
%same width broadening over the energy bands for the same pulses (see section 4.2).
%In the end we report various correlations between the model parameters
%with isotropic peak luminosity ($L_{iso}$) and isotropic energy ($E_{\gamma,iso}$)
%and discuss their implications (section 4.3). 

\section{GRB 090618}

The brightest Gamma Ray Burst (GRB) in the \emph{Fermi} era (as of March
2010), GRB 090618, 
 was simultaneously observed with \emph{Swift}/BAT
(Schady et al. 2009a; 2009b; 2009c),\emph{ AGILE} (Longo et al. 2009),
\emph{Fermi}/GBM (McBreen et al. 2009), \emph{Suzaku}/WAM (Kono et
al.), \emph{Wind}/Konus-Wind, \emph{Coronas-Photon}/Konus-RF (Golenetskii
et al. 2009), \emph{Coronas-Photon}/RT-2 (Rao et al. 2009; 2011). This is
a long GRB (T$_{90}$$\sim$113 s) with a fluence of $3398.1\pm62.0\times10^{-7}$
erg/cm$^{2}$ (integrated over $\triangle t$=182.27 s).
The detection time reported by \emph{Swift}/BAT is  2009 June 18 at 08:28:29.85 UT (Schady et al.
2009a) whereas  the \emph{Fermi}/GBM detection time is 08:28:26.66
UT (McBreen et al. 2009). This burst showed structures in the light
curve with four peaks along with a precursor. A significant
spectral evolution was also observed in the prompt emission 
which lasted for about 130 s from the trigger time (for
a summary see Rao et al. 2011).

The GRB afterglow was observed in the softer
X-rays and optical regions. The optical afterglow was tracked by Palomar
60-inch telescope (Cenko 2009), Katzman Automatic Imaging
Telescope (Perley et al. 2009) followed by various other optical,
infrared, and radio observations. Observations using the Kast spectrograph on the 3-m Shane
telescope at Lick Observatory (Cenko et al. 2009) revealed a redshift of z = 0.54 (Cenko
et al. 2009a). The X-ray afterglow was measured by the \emph{Swift}/XRT (Schady
et al. 2009b). Initially very bright in the X-rays, the flux rapidly
decayed with a slope of $\sim$6 before breaking after 310 s from the
trigger time when the burst entered a shallower decay phase (slope
0.71$\pm$0.02 -- Beardmore et al. 2009). 

Rao et al. (2011) analyzed the prompt emission of GRB 090618 using
\emph{Coronas-Photon}/RT-2 and \emph{Swift}/BAT data. The light curves
in various energy bands were fitted with a FRED profile to identify
pulse characteristics such as width, peak position etc. A joint RT-2
and BAT spectral analysis of individual pulses was done using the spectral
model of Band. The reported time integrated low energy photon index $\left(\alpha\right)$,
the high energy photon index $\left(\beta\right)$, and the peak energy
$\left(E_{peak}\right)$ are -1.40, -2.50, and 164 keV, respectively.
Ghirlanda et al. (2010) performed a time resolved spectral analysis
of \emph{Fermi} GRBs of known redshifts (GRB 090618 being one of the twelve
in the list) to show that the time-resolved correlation $\left(E_{peak}^{t}-L_{iso}^{t}\right)$
within individual GRBs is consistent with time-integrated correlation
$\left(E_{peak}-L_{iso}\right)$ indicating that there
must be a physical origin for the correlation and may not be an outcome of instrumental selection effects.

\section{DATA ANALYSIS}

We used \emph{Fermi}/GBM and \emph{Swift}/BAT data for our analysis. 
The GBM consists of twelve Sodium Iodide (NaI: nx where x
runs from 0 to 11) and two Bismuth Germanate (BGO : by, where y runs
from 0 to 1) detectors. The NaI covers the energy range $\gtrsim$
8 keV to $\sim$ 1 MeV while the BGO covers a wider range of $\sim$ 200 keV
to $\sim$ 40 MeV. Data are saved in three types of packets (Meegan et. al. 2009): (1) TTE
(Time Tag Events) in which time and energy information (in 128 energy channels)
 of individual
photons are  stored, (2) CSPEC: binned  data in
1.024 s bins and 128 energy channels and (3) CTIME: binned data in
0.064 s bins and 8 energy channels (which is not suitable for spectroscopy).
The triggered detectors for the GRB 090618
were n4, n7, b0, and b1. In the following analysis we have used n4,
n7 and either b0 or  b1, one at a time. On the other hand, BAT
has an array of CdZnTe detector modules located behind a Coded Aperture
Mask. This covers an energy range of 15 keV to 150 keV (Barthelmy
et al. 2005; Gehrels et al. 2004). The large area of BAT (5200 $\rm cm^{2}$)
gives very good sensitivity for timing analysis, compared to that
of NaI/BGO (126 $\rm cm^{2}$) detectors of GBM.

\subsection{Spectral Analysis}

%To start with, we wanted to quantify the utility of a joint BAT and
%GBM analysis. To compare with the independent GBM results of 
Ghirlanda et al. (2010) have presented the  time resolved spectral analysis results
for GRB~090618 in 14 time segments. We essentially followed the same methodology,
but have also used the simultaneous \emph{Swift}/BAT data. 
%We used n4 and n7 of Fermi/GBM
%along with Swift/BAT to measure the spectral parameters for a cut-off
%power law  fit. 
%The derived parameter values (using GBM alone) were found to be
%consistent with the values presented in 
%   checked by comparing
%with those reported by 
%Ghirlanda et al. (2010).
%   (using GBM alone) and then
%The parameters were redetermined using a joint analysis of the Fermi/GBM
%and the Swift/BAT data.
%
GRB~090618 triggered the \emph{Swift}/BAT instrument on 2009-06-18
at 08:28:29.851 or 267006514.688 MET (s).
The \emph{Fermi}/GBM trigger time is $T_{0}$(GBM) = 2009-06-18 08:28:26.659
UT or 267006508.659 MET (s). Hence there is a delay time in the BAT
data file which must be subtracted during the joint analysis. The
calculated time delay is -3.192 s and when we convert it to MET it gives
267006511.496 MET (s) for \emph{Swift}/BAT. In all the analysis this
time is subtracted from the trigger time unless specified otherwise.

For the GBM we chose the TTE (Time Tag Event) files of the triggered
NaI detectors, namely n4 and n7. The BGO detector was excluded as it
is not suitable for time resolved spectroscopy due to its lower sensitivity
(Ghirlanda et al. 2010). The energy range was chosen to be $\gtrsim$
8 keV to $\sim$ 900 keV. We used \emph{RMFIT} package provided by the 
\begin{figure*}
 
\centering
\includegraphics[angle=0,scale=0.3]{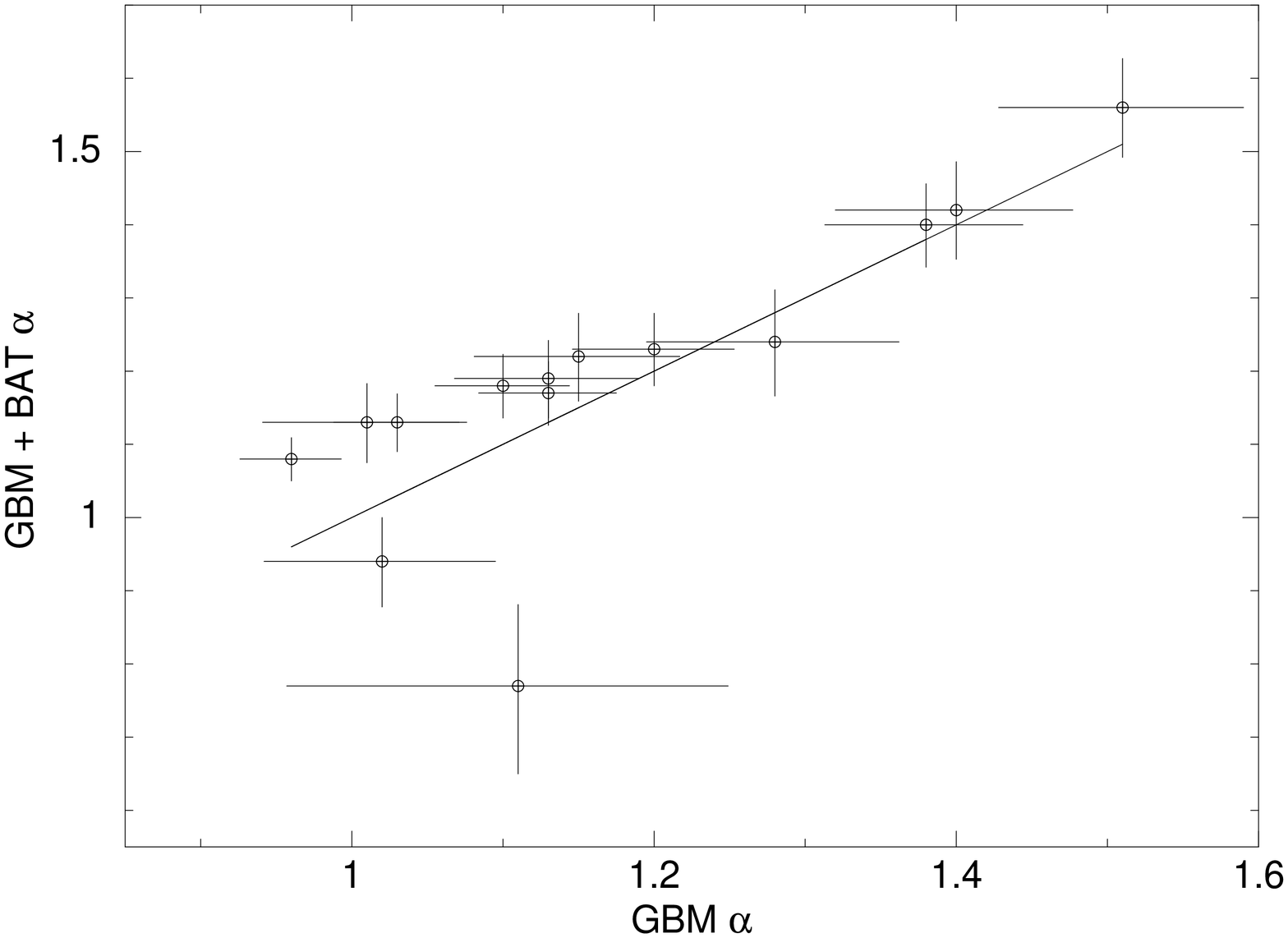}
\includegraphics[angle=0,scale=0.3]{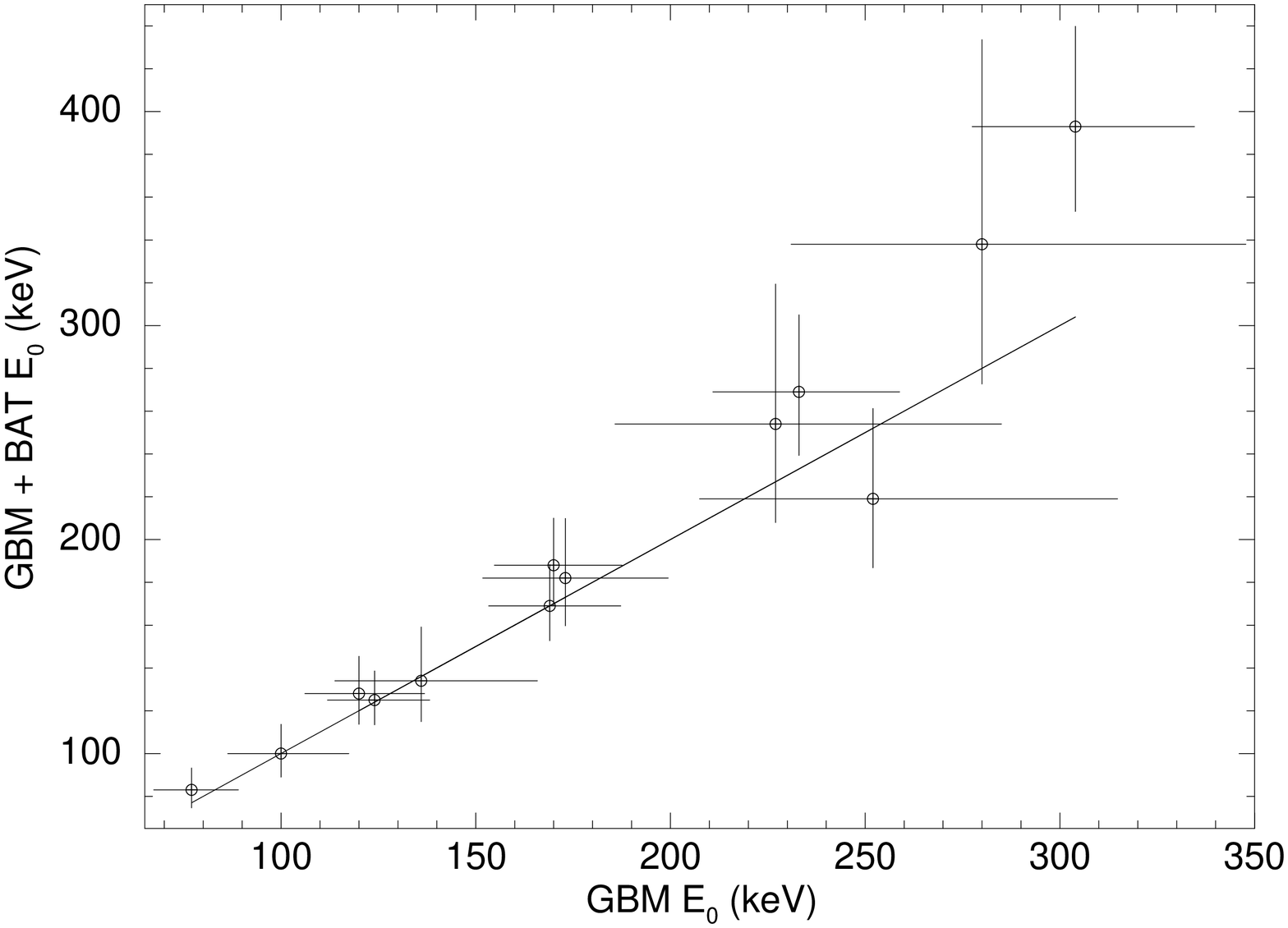}

\caption{
Comparison of the  parameters obtained from  a time resolved spectral analysis of
the prompt emission of 
GRB~090618 using a cut-off power law model. The left panel shows
the power law indices  obtained from a joint GBM and  BAT analysis  plotted
against those  obtained from a GBM analysis. The right panel shows 
a similar plot for the
cutoff energy $E_{0}$. In both cases the straight line has a slope 
1 denoting the equality of measurements ($-\alpha$ is plotted 
for convenience).
}

\end{figure*}
User Contributions
of Fermi Science Support Center (\emph{FSSC}) to select the desired time
cuts and for generating the background files. The background was taken
before and after the burst, avoiding burst contamination (Ghirlanda et al. 2010)
and the exposure time for background was optimized for different detectors.
The energy channels were binned so as to get a minimum of 40 counts
per channel. For the analysis of the BAT data we used
standard tasks prescribed in \emph{The Swift BAT Software Guide,
Version 6.3} of \emph{Heasoft 6.9} employing a time delay of -3.192
s. 

These spectra were then fitted with a cut-off power law of the form
$F=cons\times E^{\alpha}exp(-E/E_{0})$ where F is the photon
flux. The constant factor, \emph{cons},  was kept free while fitting essentially to take care of the
different systematic errors in the area calibration of the two
instruments. 
The derived parameter values (using GBM alone) were found to be
consistent with the values presented in Ghirlanda et al. (2010).
Figure 1 shows the comparison results of the
spectral analysis. In Figure 1 (left panel) the indices $\alpha$ obtained from the
joint analysis are plotted against those obtained from an analysis of the GBM data
for the fourteen different time segments. Similar data for the parameter $E_{0}$
are shown in the right panel. A deviation is noted for the parameter $\alpha$
in the 0 to 3 s region. This may be attributed to lower count rate
of BAT in the precursor region.

We note a systematic increment in the value of $\alpha$ and $E_{0}$ after the inclusion
of BAT data. Further, the constant factor is always lower by 10-20\% for
the BAT. Sakamoto et al. (2011) have made a joint spectral analysis
of 14 GRBs using data from \emph{Swift}/BAT, \emph{Konus}/Wind and \emph{Suzaku}/WAM and
they have noted that for the BAT data, the constant factor is systematically
lower by 10 \textendash{} 20\%, the photon index is steeper by 0.1
\textendash{} 0.2 and the $E_{peak}$ becomes higher by 10 \textendash{}
20\%. The results presented here are in general agreement with these
conclusions. 

In Figure 2 we have plotted the ratio of fractional errors
as a function of the measured parameters. We find that there is a
marginal improvement for the parameter $E_{0}$ (Figure 2, right panel). 
The parameter $\alpha$ (Figure 2, left panel),
however is measured always with a greater accuracy from the joint analysis.
The large effective area of BAT improves the accuracy of the spectral
index whereas its low high energy threshold
($\sim$150 keV) makes relatively low impact on the measurement
of the high energy cut-off value. The joint analysis always gives
confidence on the measured parameters and ensures that the effects
of instrumental artefacts are minimized. Hence we use, in the later
analysis, a joint fit for measuring the global spec-
\begin{figure*}\centering
\includegraphics[angle=0,scale=0.3]{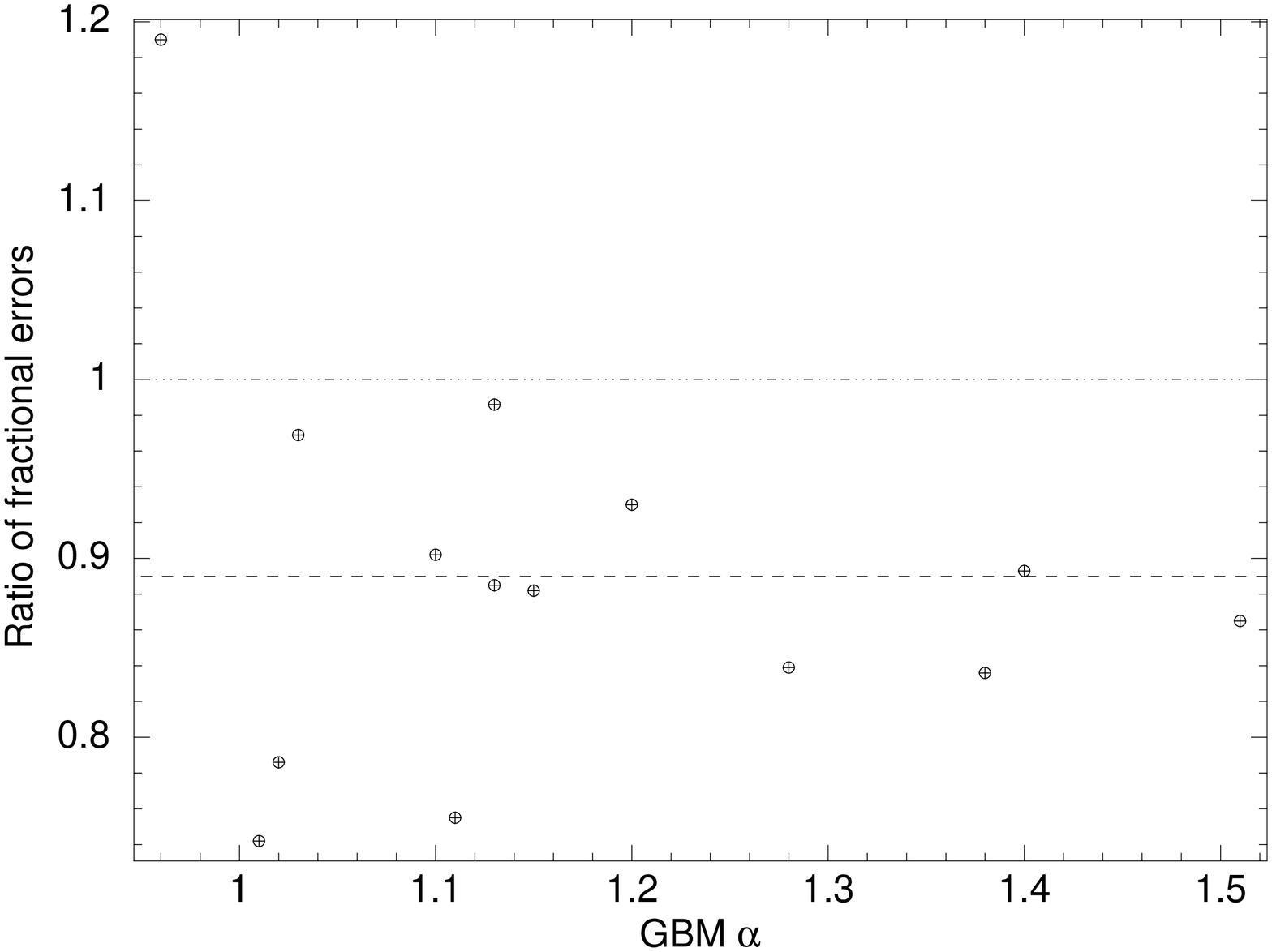}
\includegraphics[angle=0,scale=0.3]{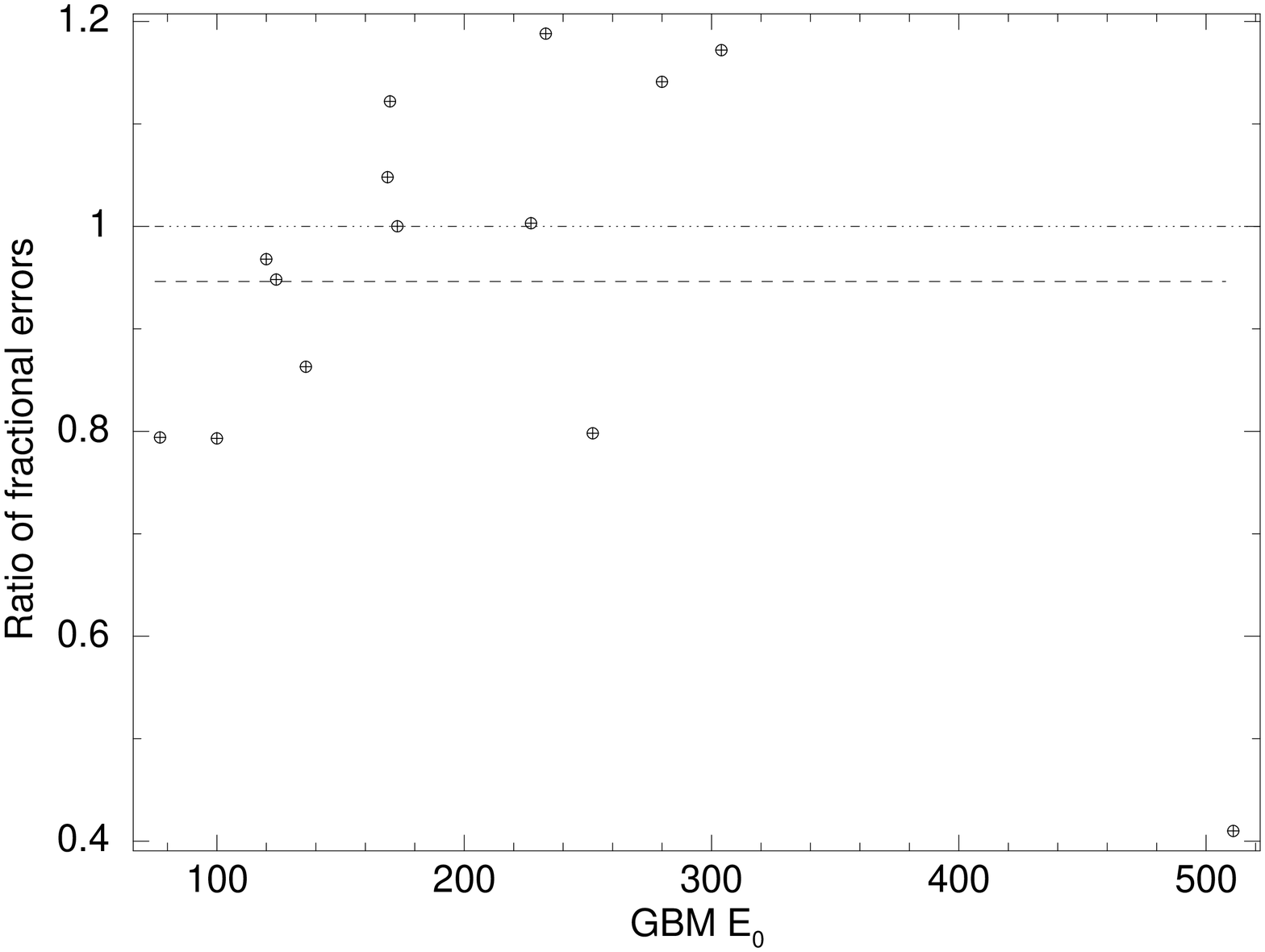}

\caption{The ratio of fractional errors for GBM-BAT joint analysis to the GBM
analysis is shown as a function of the measured values of $\alpha$
(left panel) and $E_{0}$ (right panel) obtained from a fitting for
GBM data alone. The dotted line in both cases shows the average of
all errors measured. This is found to be marginally less than the line
of equal error (dot dashed) for the $E_{0}$ and well below that line
for $\alpha$ measurement ($-\alpha$ is plotted
for convenience).}

\end{figure*}
tral parameters
and use only the BAT data for a joint timing and spectral analysis.

\subsection{Timing Analysis}

We identified four pulses (other than the precursor) in the $\sim$60
s to $\sim$130 s period after the trigger. Both the \emph{Swift}/BAT and
\emph{Fermi}/GBM (n4) were used for the timing analysis. We subtracted background
and shifted the time origin to the trigger time (TRIGTIME is the keyword 
in the header) of the respective satellite. As we noted previously,
there is a time lag of -3.192 s for the  BAT trigger.
Light curves were extracted for different energy bands for both
the data sets. The energy bands for the \emph{Swift}/BAT were:
15-25 keV, 25-50 keV, 50-100 keV, and 100-200 keV. For the \emph{Fermi}/GBM
(n4) we chose 8-15 keV, 15-25 keV, 25-50 keV, 50-100 keV, 100-200
keV, and 200-500 keV. Greater than 500 keV region of NaI and the BGO
detector of GBM were excluded because the third and the fourth pulses
were not detected in these bands.

All the light curves were fitted with the four-parameter model of
Norris et al. (2005)
\begin{equation}
I(t)=A_n\lambda exp\{-\tau_{1}/(t-t_{s})-(t-t_{s})/\tau_{2}\}\label{1}\end{equation}
for $t>t_{s}$, where $\mu=\left(\tau_{1}/\tau_{2}\right)^{\frac{1}{2}}$ and
$\lambda=exp\left(2\mu\right)$. $A_n$ is the pulse amplitude, $t_{s}$
is the pulse start time and $\tau_{1}$, $\tau_{2}$ are time constants
characterizing the rise and decay part of the pulse. From these parameters, 
various derived quantities like the peak position, the pulse width 
(measured as the separation between the two 1/e intensity points), and the
pulse asymmetry   can be
calculated 
%from parameter estimates for example, the peak position of the pulse ($\tau_{peak}$),
%the width measured between the two 1/e intensity points ($w$) and the pulse
%asymmetry ($\kappa$) 
(see Norris et al. 2005 for details). 
Errors are calculated for the parameters as nominal 90\% confidence level errors
($\triangle\chi^{2}$=2.7) and these errors are propagated for the
derived quantities, as described in
Norris et al. (2005).

Figure 3 shows the pulse width (w) as a function of energy (E) for all the pulses.
Apart from the fact that the width broadens in the lower energy bands,
we note a tentative evidence for  a decrement (or constancy within error bar) of the
width from 50-100 keV through 8-15 keV energy bands for the third and
the fourth pulses (lower panels of Figure 3). We fitted these data for the individual pulses
with (i) constant ($w=c$) and (ii) linear ($w=mE+c$) models.
For the first two pulses there was a significant improvement in the  $\chi^2$ 
for a linear fit (the $\chi^2$ varied from 16.2(9)
to 5.1(8) and 140.9(9) to 7.9(8) respectively for the two pulses; 
the numbers in the brackets are the degrees of freedom). 
The derived slopes are $(-7.4\pm2.6)\times10^{-3}$ s $\rm keV^{-1}$ and
$(-5.5\pm0.4)\times10^{-3}$ s $\rm keV^{-1}$, respectively 
for the two pulses. The third and
the fourth pulses, however,  showed negligible improvement in 
\begin{figure}[H]\centering
\includegraphics[angle=0,scale=0.3]{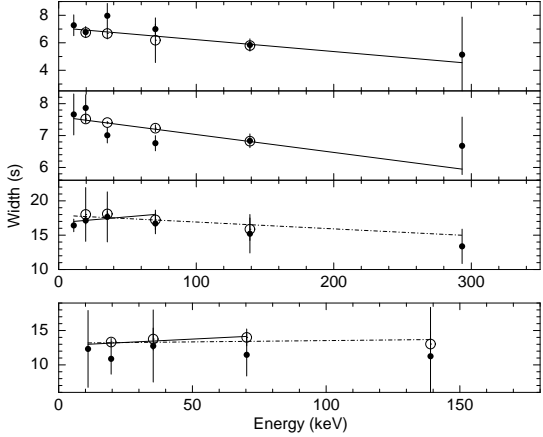}\caption{Pulse width (w) is plotted against energy (E),
for pulse 1 to 4 respectively in the top to bottom panels. The fourth pulse is insignificant in the
200-500 keV energy range. Hence the fourth panel has a different scale on the x-axis.Values derived from
the Swift/BAT  and the Fermi/GBM data are shown by circles and filled circles, respectively. While the width
decreases with increasing energy for the first two pulses (top two panels), there is a tentative evidence
of the reverse effect in the lower energy ($\sim$ 10-70 keV band) for the other two pulses (see section 3.2).}

\end{figure}
$\chi^2$ (from 8.7(9)
to 6.0(8) and 2.1(8) to 2.1(7),  respectively for the two pulses) for the linear fit.  
For these two pulses, when the data below 70 keV is considered, there 
is a marginal improvement in the $\chi^2$ for a linear fit with a positive
slope (from 8.7(9) to 
%Linear fit in the lower energy band ($\sim$ 10 - 70 keV), however, gives
3.8(5) and 2.1(8) to  1.5(5),  respectively for the two pulses). The 
derived slopes are 
$(15\pm14)\times10^{-3}$ s $\rm keV^{-1}$ and $(10\pm21)\times10^{-3}$ s $\rm keV^{-1}$
respectively.
Though the slopes for these two pulses are positive, the error bars are quite large
and hence the evidence of this reverse variation is \emph{tentative}. In Figure 3 (lower two panels)
the linear fit for pulse 3 and 4 are shown by solid lines (in the low energy) and 
dot-dashed lines (in full energy band). We have ignored the fourth pulse 
in the 200-500 keV band of Fermi for the fitting as it was not prominent in this energy band.

\subsection{Pulse-Wise Joint Spectral Analysis}

The peak position and the width gave a clue for the time cut for each
pulse and facilitated the selection of integration time for pulse
spectral analysis of the individual pulses. The pulse spectral analysis
was done with a Band function fit and including the BGO detector of
GBM. We included the precursor pulse in this analysis. We divided
the GRB event into four time segments namely, Part 1 ($T_{0}$ to
$T_{0}+50$), Part 2 ($T_{0}+50$ to $T_{0}+77$), Part 3 ($T_{0}+77$
to $T_{0}+100$) and Part 4 ($T_{0}+100$ to $T_{0}+180$). 
The first and second pulses are considered together as there is too much
overlap between them.

All the spectra  were fitted in the energy range $\sim$ 8 keV - $\sim$
1 MeV with the four parameter model of Band (Band et al. 1993):

\begin{equation}
I(E) = \left\{ \begin{array}{ll}
 A_{b}\left[\frac{E}{100}\right]^{\alpha}exp\left[\frac{-(2+\alpha)E}{E_{peak}}\right] \\
\\    
$\rm if$ ~~E\le [(\alpha-\beta)/(2+\alpha)]E_{peak}\\
\\
  A_{b}\left[\frac{E}{100}\right]^{\beta}exp\left[\beta-\alpha\right]\left[\frac{\left(\alpha-\beta\right)E_{peak}}
     {100\left(2+\alpha\right)}\right]^{(\alpha-\beta)} \\
\\
otherwise
       \end{array} \right.
\end{equation}

The lower and the higher
energy parts are characterized by the parameters $\alpha$ and $\beta$
respectively and they meet smoothly at the spectral break energy $[(\alpha-\beta)/(2+\alpha)]E_{peak}$.
The peak energy, $E_{peak}$ or equivalently the
characteristic energy, defined as $E_{0}=E_{peak}/\left(2+\alpha\right)$, 
may be considered as the third parameter of the model. The best fit model parameters along with
the nominal 90\% confidence level error ($\triangle\chi^{2}$=2.7) for the joint
fit of \emph{Fermi}/GBM (n4, n7, b1) and \emph{Swift}/BAT are reported
in Table 1 (b0 in place of b1 gives similar results). 
The high energy photon index $\beta$ is constrained primarily by the GBM data and hence we have
kept its value fixed at -2.5 to determine precise values of $E_{peak}$ and $\alpha$. 
Only in one case namely for the full range ($T_{0}$ to $T_{0}+180$ sec) of the GRB, 
$\beta$ was made to vary and a value of -2.96 was obtained.
Figure 4 shows the joint Band spectral fitting of this range along with the $\chi^{2}$ deviation of the data.

An examination of the residuals in Figure 4 reveals that there are
some systematic differences in the response in the 20-40 keV 
region between BAT and GBM data and the resultant $\chi^{2}$
\begin{table*}\centering
\emph{\caption{spectral parameters of GRB 090618 for Band model,
 from a joint analysis of \emph{Fermi}/GBM (n4,
n7, b1) and \emph{Swift}/BAT data}
}

\begin{tabular}{>{\centering}p{0.5in}ccccc}
\hline 
\multicolumn{1}{>{\centering}p{1.5in}}{Part} & $\alpha$ & $\beta$ & $E_{peak}$(keV) & $\chi^{2}$(dof) & $E_{0}$ (keV)\tabularnewline
\hline
\hline 
\multicolumn{1}{>{\centering}p{1.5in}}{Full} & -1.35$\pm0.02$ & -2.5 (fixed) & 154$\pm7$ & 428(299) & 236$\pm18$\tabularnewline
\multicolumn{1}{>{\centering}p{1.5in}}{Part 1 (Precursor)} & \multicolumn{1}{c}{-1.25$\pm0.05$} & -2.5(fixed) & $166_{-14}^{+18}$ & 369(299) & $221_{-33}^{+39}$\tabularnewline
\multicolumn{1}{>{\centering}p{1.5in}}{Part 2 (Pulse 1 \& 2)} & -1.11$\pm0.02$ & -2.5(fixed) & 212$\pm9$ & 460(299) & 238$\pm15$\tabularnewline
\multicolumn{1}{>{\centering}p{1.5in}}{Part 3 (Pulse 3)} & -1.15$\pm0.03$ & -2.5(fixed) & 114$\pm4$ & 334(259) & 134$\pm9$\tabularnewline
\multicolumn{1}{>{\centering}p{1.5in}}{Part 4 (Pulse 4)} & $-1.65_{-0.08}^{+0.11}$ & -2.5(fixed) & 33$\pm3$ & 431(299) & $94_{-30}^{+38}$\tabularnewline
\multicolumn{1}{>{\centering}p{1.5in}}{Full (free $\beta$)} & -1.36$\pm0.02$ & -2.96$\pm$0.48 & $160_{-8}^{+9}$ & 418(288) & $250_{-20}^{+22}$\tabularnewline
\hline
\end{tabular}
\end{table*}

\begin{figure}[H]\centering
\includegraphics[angle=0,scale=0.3]{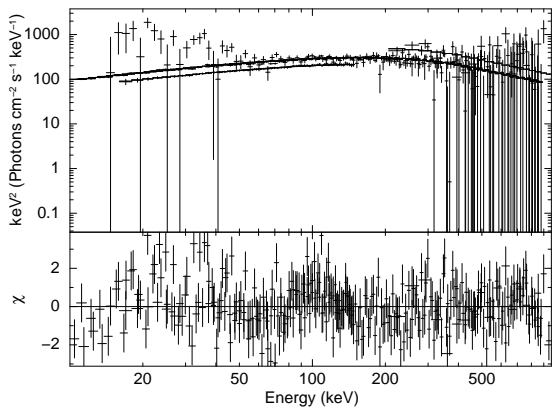}\caption{Fitting of Band function to the full range (0 to 180 s) of GRB 090618
for the  energy range $\sim$ 8 keV to $\sim$ 1 MeV. The range of the
BGO detector starts above 200 keV, while BAT covers up to $\sim$ 150 keV.
The other two detectors namely n4 and n7 have the spectrum above 8 
keV to $\sim$ 800 keV. The extreme channels and specifically the
25th channel of n4 are ignored due to high background rates. The unfolded
spectrum of BAT shows a lower normalization (see text).}

\end{figure}

values generally are quite high. The resultant spectral parameters,
however, agree with the results reported for this GRB (Ghirlanda et
al., 2010; Nava et al. 2011; Rao et al. 2011). The errors are significantly
better compared to the joint BAT and RT-2 analysis (Rao et al. 2011),
but comparable to that obtained from a GBM analysis alone (Nava et al.
2011). Since a joint analysis using data from different satellites pins down
the systematic errors in each instruments, we believe that the joint
analysis results, reported here, are devoid of instrumental artefacts,
particularly for the parameters $\alpha$ and $E_{peak}$.

\section{JOINT SPECTRAL AND TIMING STUDY: AN ALTERNATE APPROACH }

\subsection{Two Parameter XSPEC Table Model }

\subsubsection{The time evolution of $E_{peak}$}
The light curve of each pulse of a GRB can be described in terms of the four-parameter model of Norris (equation 1) and the spectrum
for each pulse can be fitted with the Band model (equation 2). Liang and Kargatis (1996) showed for clean,
separable pulses that the peak energy
of the $\nu F_{\nu}$ spectrum, $E_{peak}$, decays with the photon
fluence as,
\begin{equation}
E_{peak}(t')=E_{peak,0} ~e^{-\phi(t')/\phi_{0}}\label{2}\end{equation}
where $E_{peak,0}$ is the peak energy at zero fluence,
$\phi_{0}$ is the decay constant and $t'=t-t_s$. The quantity $\phi(t')$ is
the fluence up to time t from the start of the pulse ($t_s$). For convenience of description
we shift our origin to $t_s$ and hence $t'=t$ for each pulse analysis. Equation 3
quantifies the overall softening of the burst with time. 

Kocevski and Liang (2003) have integrated the light curves of a sample of GRBs
up to various times t and found the fluence, $\phi(t)$, at those times. The
peak energy at the same times, $E_{peak}(t)$, were found by fitting the spectra with a 
Band model. They determined the $E_{peak,0}$ and $\phi_{0}$ from
a semi-log plot of $E_{peak}(t)$ versus $\phi(t)$ and fitting a straight
line. This procedure was adopted for clean, bright, and separable
FRED pulses.

%GRB 090618 shows multiple overlapping pulse structure
%and hence it is not recommended to integrate the actual light curve
%to get $\phi(t)$.
%Therefore 
Here we explore a method to determine the timing and spectral parameters by \emph{assuming
that $E_{peak}(t)$ and $\phi(t)$ are related according to equation 3 in each pulse.}
Essentially we assume that for each pulse, the variation of flux with time is given 
by the exponential model (equation 1), and at each time the spectra is described by 
a Band function (equation 2), and, for a given pulse, only the peak energy 
varies with time governed by equation 3. We develop a method to derive the model
parameters from a  time integrated spectrum and energy integrated light curve.

\subsubsection{Description of the method}

The flux of a GRB pulse is a function of time and energy. If there exist
empirical formulae for the temporal and spectral beahvior then the flux can 
be described in terms of all the model parameters. For example,
the variation of flux with time is given by Norris\textquoteright{}
exponential model, with four  parameters ($A_{n}$, the normalization,
$t_s$, $\tau_{1}$,  and $\tau_{2}$). The spectral behaviour is parametrized
by a Band function, with four  parameters ($A_{b}$, the normalization,
$\alpha$, $\beta$, and $E_{peak}$). For any timing model, the time parameters change with energy
e.g., the derived parameter, width, changes in various energy bands (see
section 3.2). Similarly, the spectral parameters of any spectral model change with time e.g., $\alpha$ and $E_0$
of cut-off power law  change with time in the same pulse (see section 3.1). 
%However, we can always find global values of some of the parameters (in that pulse, not
%for the full GRB) e.g., $\alpha$ and $\beta$
%of each pulse (see section 3.3), and let the others vary.

%For clarity let us now consider in particular t
The Norris exponential model (equation 1)  
describes the evolution of a   pulse with time. This description
can be used for counts in any energy band and the variation of the
model parameters with energy can be studied. Instead, 
% at and the spectral model
%of Band for a single pulse (see equations (1) and (2)). These are the models 
%we have chosen for our mehod. 
we assume that the model parameters $\tau_{1}$  and $\tau_{2}$ derived for a pulse
integrated over all energy bands are 
global parameters describing the evolution of total flux with time.
Since the variation of spectrum is best described by the peak energy evolution
with fluence (equation 3), 
we further assume that $\alpha$
and $\beta$ do not change with time.
% and $E_{peak}(t)$
%varies with the fluence from the start of the pulse, as given by equation (3). 
The normalization factors 
($A_{n}$ and $A_{b}$) are related by the fact that the fluence calculated either
by time integration of the pulse light curve or by the integration over the total energy 
should be the same. With these assumptions, the parameters for a given pulse are determined as follows.
\begin{table*}\centering
\caption{Best fit values of $E_{peak,0}$ and $\phi_{0}$ for the pulses
as determined by the $\chi^{2}$ minimization of the spectral data
with the additive table model. The first two pulses are combined together.
The other pulse characteristics ($\tau_{1}$,
$\tau_{2}$, and $t_{s}$) are also reported.}

\begin{tabular}{>{\centering}p{0.45in}>{\centering}p{0.5in}>{\centering}p{0.6in}>{\centering}p{0.75in}>{\centering}p{0.8in}>{\centering}p{0.7in}>{\centering}p{0.7in}>{\centering}p{0.7in}}
\hline 
Pulse & $E_{peak,0}$ & $\phi_{0}$ & Norm & $\chi^{2}$ (dof) & $\tau_{1}$(s) & $\tau_{2}$(s) & $t_{s}$(s)\tabularnewline
\hline
\hline 
1 & $359_{-92}^{+65}$ & $12.2_{-1.3}^{+2.5}$ & $0.74\pm0.06$ & 40.12 (75) & $795.4_{-7.1}^{+7.2}$ & $0.54_{-0.005}^{+0.005}$ & \multicolumn{1}{>{\centering}p{0.7in}}{$40.1_{-0.2}^{+0.2}$}\tabularnewline
\hline 
2 &  &  &  &  & $858.2_{-7.7}^{+6.4}$ & $0.58_{-0.005}^{+0.004}$ & $44.2_{-0.2}^{+0.2}$\tabularnewline
\hline 
3 & $324_{-83}^{+82}$ & $18.8_{-6.2}^{+9.6}$ & $0.54\pm0.04$ & 31.41 (75) & $353.4_{-50}^{+43}$ & \multicolumn{1}{>{\centering}p{0.7in}}{$2.47_{-0.09}^{+0.12}$} & $50.3_{-1.2}^{+1.5}$\tabularnewline
\hline 
4 & $307_{-99}^{+41}$ & $12.0_{-2.1}^{+2.0}$ & $0.19\pm0.04$ & 56.42 (75) & $532.0_{-132}^{+209}$ & \multicolumn{1}{>{\centering}p{0.7in}}{$1.58_{-0.16}^{+0.15}$} & \multicolumn{1}{>{\centering}p{0.7in}}{$81.0_{-3.4}^{+2.6}$}\tabularnewline
\hline
\end{tabular}
\end{table*}

For a given pulse, we use the energy integrated light curve from BAT and use equation 1
for the fitting. This best fit light curve is
used to determine the global parameters namely, $A_{n}$, $\tau_{1}$, and $\tau_{2}$. 
Since we can arbitrarily choose the pulse starting time by shifting the time origin,
the parameter $t_s$ is not very relevant (also see section 4.1.1). 
This function is then integrated up to a time t to determine $\phi(t)$. From the
individual spectral fitting for each pulse using the Band model, $\alpha$
and $\beta$ are determined (see section 3.3). Now, if the constant
parameters namely, $E_{peak,0}$ and $\phi_{0}$
are determined, we have a complete spectral and timing description
of a   GRB pulse. For this purpose, we use a two-dimensional grid 
of guess values for these two parameters and then determine the best fit values
by a spectral fitting method.

For a particular combination of $E_{peak,0}$ and $\phi_{0}$
values, $E_{peak}(t)$ at any time t can be determined from equation 3. 
On the other hand, equation 2 can be rewritten as $I(E)=A_b~f(E)$ where $A_b$
is the normalization and f(E) is the rest of the functional form valid in
the two energy bands. Given the global parameters ($\alpha$ and $\beta$) and
the time dependent parameter ($E_{peak}(t)$), f(E) can be determined. 
The normalization $A_{b}(t)$ can be found by dividing the fluence ($\phi$(t)) of the light curve
by the f(E) integrated over the whole energy range. Hence photon spectrum is 
generated for the time instance t. Choosing a time and energy grid, counts can be plotted as a function of time
and energy for a particular combination of $E_{peak,0}$ and $\phi_{0}$ values, and the resulting surface can be
referred to as a three dimensional pulse model. Hence for a two-dimensional grid
of guess values of $E_{peak,0}$ and $\phi_{0}$ we can generate three dimensional
pulse models for each of the grid points. These model pulses can be used in 
two ways: (1) integration over time gives the model spectrum of that pulse, 
(2) integration over various energy bands gives model light curves of that 
pulse in those energy bands. 

In our analysis, we first generate a two parameter XSPEC table model using the guess 
values of $E_{peak,0}$ and $\phi_{0}$ and then 
use this table model to determine the best fit values of these two parameters by fitting
the time integrated spectrum 
(see section 4.1.3). With this method, we get a complete description of the pulse with
model parameters determined from the data.

\begin{figure}[H]\centering
\includegraphics[scale=0.50]{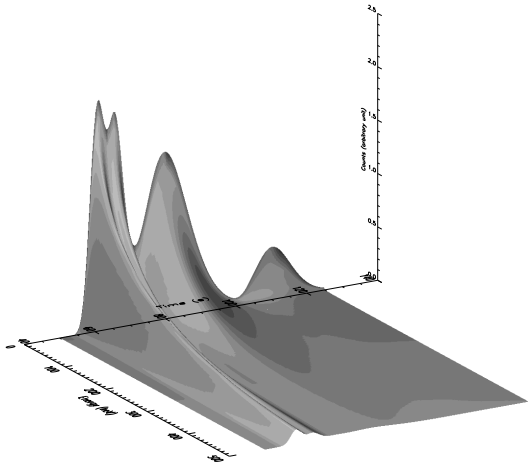}\caption{The three dimensional model of the whole GRB using
a model with the best fit values of $E_{peak,0}$ and $\phi_0$ for each pulse.
Counts in arbitrary unit is plotted in the vertical axis against
time and energy. The pulses are shifted to original pulse start time (see Table 2).
}

\end{figure}
Using this model we can reconstruct the three dimensional 
pulse profiles and can generate 
light curves in any energy band. Various timing parameters 
(width and lag) are derived in these energy bands (see section 4.2)  and these
model predicted values can be compared with the observed values to get a validation of
the whole procedure.

\subsubsection{Specification of the table model and the best fit values}
For all the pulses we use a time grid resolution of 0.5 s and energy grid resolution of 2.0
keV. Each pulse has a separate XSPEC table model, though the first and second pulses are
combined together as the data overlap. A typical table
model contains a spectrum in the 2-200 keV energy range for 200 values of
$E_{peak,0}$ from 100.0 keV to 1100.0 keV and 50 values of $\phi_{0}$
from 2.0 photon $\rm cm^{-2}$ to 77.0 photon $\rm cm^{-2}$, energy binning
information, parameter values and the required header keywords standardized
as per the requirement by XSPEC described in OGIP Memo OGIP/92-009. 

The additive table model is used to determine the best fit values
of $E_{peak,0}$ and $\phi_{0}$ by the standard $\chi^{2}$ minimization
procedure. The best fit values with nominal 90\% confidence errors are reported in Table 2. 
The parameters, $E_{peak,0}$ and $\phi_{0}$ of the first two pulses are
assumed to be the same. It should be noted here that we have used
the normalization $A_n$ as a global parameter to calculate $\phi$(t). 
The normalization of the subsequent pulse profile is derived from the 
integration of f(E) over the total energy band (see section 4.1.2). 
Effectively this leads to a spectrum with normalization 
\begin{figure*}\centering

\subfloat{\includegraphics[angle=0,scale=0.3]{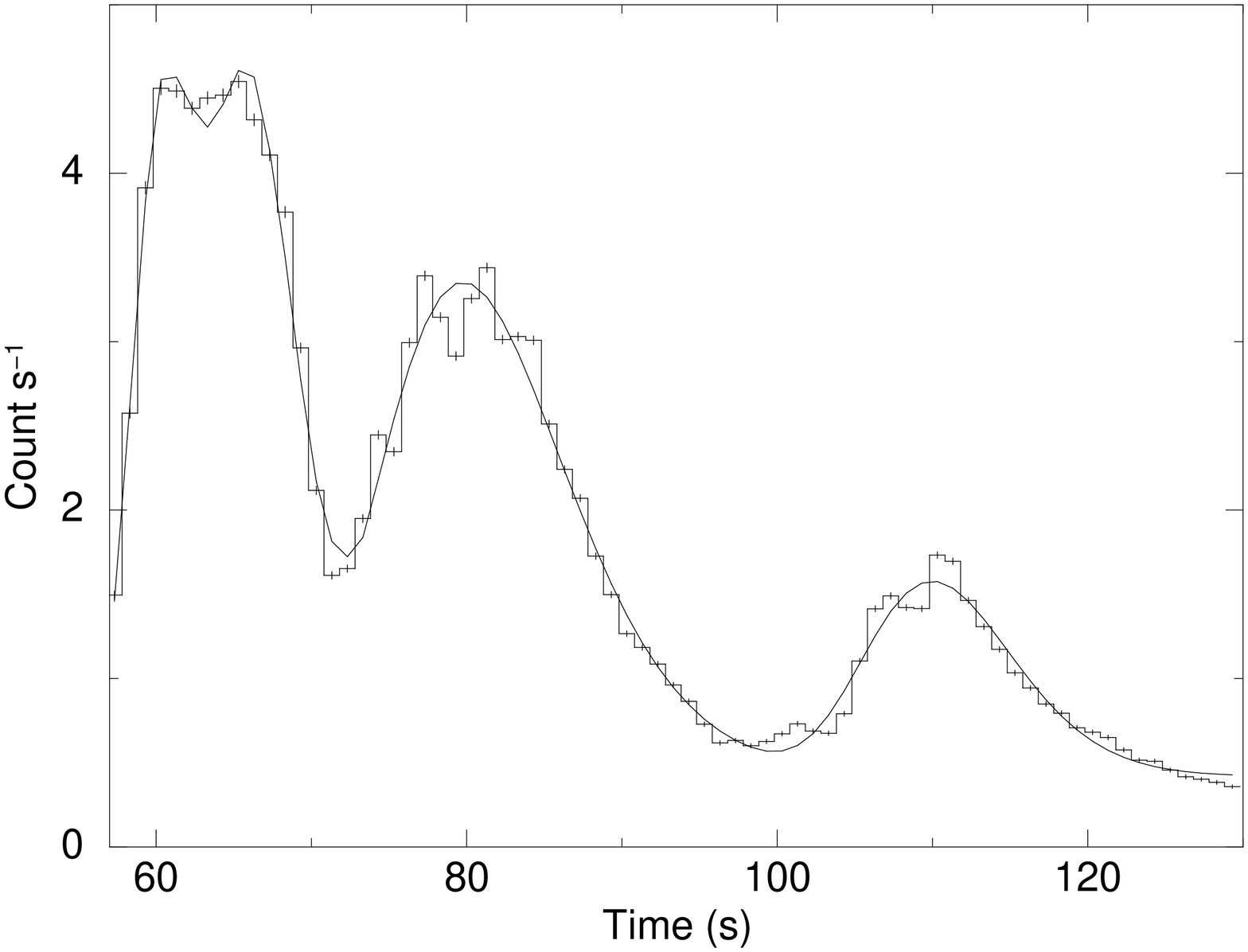}}\subfloat{\includegraphics[angle=0,scale=0.45]{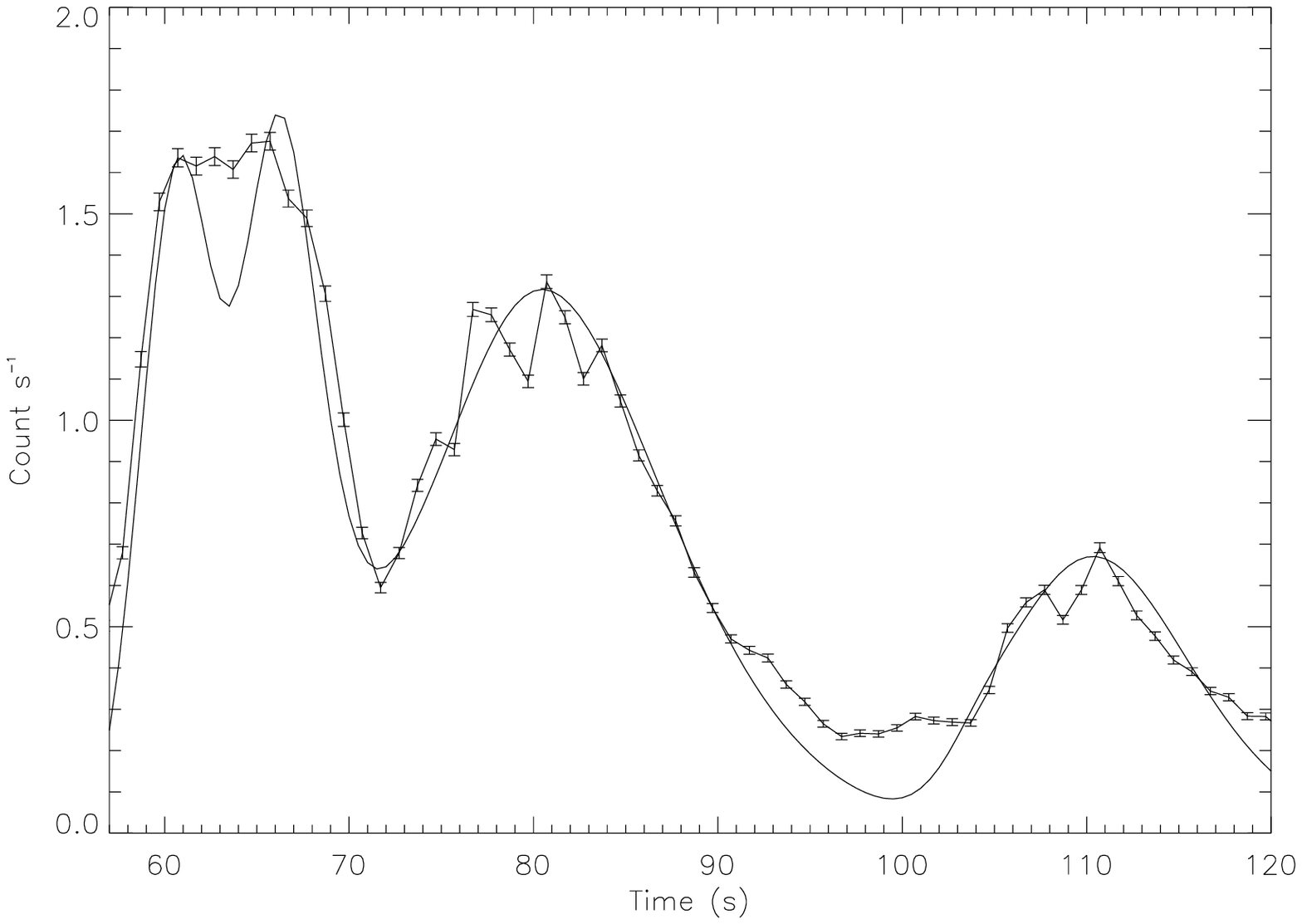}

}\caption{Observed light curve in the full energy range 15-200 keV 
fitted with the Norris' exponential model is shown in the left panel. The timing
parameters are determined and used in the pulse generation to predict light
curves in all energy bands. One such predicted light curve in the 25-50 
keV region is shown along with the observed light curve in the right panel. The 
count/s in the y-axis is BAT  mask weighted count $\rm s{-1}$.}

\end{figure*}
as a free parameter. 
This normalization is determined during the spectral fitting using XSPEC, and
is also reported in Table 2.

\subsubsection{Three dimensional pulse model for the best fit values}
Three dimensional model of each pulse is generated for the 
best fit parameter values (as described in sections 4.1.2 and 4.1.3).
The three dimensional model of the whole GRB obtained from the sum of 
these models of the pulses with each pulse shifted to its
proper starting time is shown in Figure 5. This figure is used only for display
purposes. In actual analysis, each pulse is normalized properly, before
calculating various parameters as the overall normalization may vary
for each pulse. This normalization is achieved for spectrum and light
curve separately. 

\subsection{Timing analysis using the model}

\subsubsection{Reproducing the light curves}
Once we get the best fit values of $E_{peak,0}$ and $\phi_{0}$ for
each pulse we can generate the best fit model light curve by integrating 
the three dimensional pulse model for the best fit value 
over the desired energy range. We generate three dimensional pulse models, shift them over
appropriate starting time, co-add them with proper normalization factors
and integrate over energy ranges to generate the combined light curves
in the energy bands: 15-25 keV, 25-50 keV, 50-100 keV and 100-200
keV of BAT. In Figure 6 (right panel), we show 
the observed light curve along with the model predicted light curve 
for the 25-50 keV energy band. In the left panel Norris' exponential model
fit for the total energy range is shown. Parameters from this fit (Figure 6, left panel) were used
for characterizing the timing parameters in our model.

The GRB light curve shows rapid time variable components over and above the smooth pulse
structure. 
In the pulse description we consider only this smooth  variation. Any
small rapid time variability in the data gives large value of $\chi^{2}$
when we use the smooth pulse model for fitting. Hence, the normalization
factor could not be determined by $\chi^{2}$ minimization. Instead
we physically examined the heights and determined the normalization.
The co-added pulses reproduce the light curve quite well.
Hence, we can conclude that the total light curve and the integrated
spectrum, with the assumption of $E_{peak}(t)$ varying with $\phi(t)$
in a predetermined way (equation (3)), correctly reproduces the energy resolved
light curves. This can be taken as a confirmation of the assumption
of the $E_{peak}$ variation.
It should, however,  be noted that 
\begin{table*}\centering
\caption{Variation of the model predicted width of each pulse with energy. Numbers
within brackets are the measured pulse widths.}

\begin{tabular}{c>{\centering}p{1in}ccc}
\hline 
Pulse & \multicolumn{1}{>{\centering}p{1in}}{15-25 keV} & \multicolumn{1}{>{\centering}p{1in}}{25-50 keV} & \multicolumn{1}{>{\centering}p{1in}}{50-100 keV} & \multicolumn{1}{>{\centering}p{1in}}{100-200 keV}\tabularnewline
\hline
\hline 
1 & $6.05_{-0.12}^{+0.12}$

($6.74_{-0.13}^{+0.13}$) & \multicolumn{1}{>{\centering}p{1in}}{$5.88_{-0.16}^{+0.16}$

($6.67_{-0.39}^{+0.39}$)} & \multicolumn{1}{>{\centering}p{1in}}{$5.54_{-0.17}^{+0.16}$

($6.18_{-1.62}^{+1.62}$)} & \multicolumn{1}{>{\centering}p{1in}}{$5.08_{-0.18}^{+0.15}$

($5.79_{-0.30}^{+0.30}$)}\tabularnewline
\hline 
2 & $6.56_{-0.19}^{+0.22}$

($7.52_{-0.13}^{+0.13}$) & \multicolumn{1}{>{\centering}p{1in}}{$6.35_{-0.19}^{+0.22}$

($7.40_{-0.04}^{+0.04}$)} & \multicolumn{1}{>{\centering}p{1in}}{$5.89_{-0.34}^{+0.27}$

($7.23_{-0.11}^{+0.11}$)} & \multicolumn{1}{>{\centering}p{1in}}{$5.33_{-0.34}^{+0.20}$

($6.82_{-0.04}^{+0.04}$)}\tabularnewline
\hline 
3 & $17.84_{-0.70}^{+0.47}$

($18.01_{-3.91}^{+3.91}$) & \multicolumn{1}{>{\centering}p{1in}}{$18.10_{-0.89}^{+0.53}$

($18.09_{-0.40}^{+0.40}$)} & \multicolumn{1}{>{\centering}p{1in}}{$18.68_{-0.96}^{+0.74}$

($17.22_{-1.43}^{+1.43}$)} & \multicolumn{1}{>{\centering}p{1in}}{$19.82_{-0.64}^{+0.49}$

($15.87_{-1.63}^{+1.63}$)}\tabularnewline
\hline 
4 & $14.45_{-0.92}^{+0.99}$

($13.32_{-0.51}^{+0.51}$) & \multicolumn{1}{>{\centering}p{1in}}{$14.66_{-0.96}^{+0.98}$

($13.76_{-1.60}^{+1.60}$)} & \multicolumn{1}{>{\centering}p{1in}}{$15.18_{-0.88}^{+0.62}$

($14.00_{-1.23}^{+1.23}$)} & \multicolumn{1}{>{\centering}p{1in}}{$15.93_{-1.27}^{+1.16}$

($13.03_{-1.84}^{+1.84}$)}\tabularnewline
\hline
\end{tabular}
\end{table*}
a single joint spectral fit is done for pulses 1 and 2
and hence the derived values of $E_{peak,0}$ and $\phi_{0}$ for the first two pulses are 
only the average values.

\subsubsection{Derived timing parameters}
Pulse width of individual pulses in the various energy bands can be derived by fitting the light curve
using equation (1) and using the best fit values of $\tau_1$ and $\tau_2$ (see section 3.2).
For overlapping pulses this method is erroneous e.g.,
rising part of one pulse can fall on top of the falling part of the previous pulse
affecting the width. Also, pulse delay characteristics are affected by
these overlaps. In such cases the method described here is very useful.
It takes as input some global parameters ($A_n$, $\tau_1$, $\tau_2$,
$\alpha$ and $\beta$) which are average quantities characterizing the 
pulse and some constant parameters ($E_{peak,0}$, $\phi_{0}$), 
determines the variable parameter ($E_{peak}(t)$) over the time grids and
generates the three dimensional model of the pulse. Note that the timing parameters $\tau_1$ and $\tau_2$
are determined from the energy integrated light curve and are fixed
in our model. The timing properties of the pulse (e.g., width in various
energy bands, spectral delay) are now determined crucially by the evolution 
of the peak energy. Integration of the three dimensional model in various
energy bands generates the light curves in those energy bands. Hence
physically we can determine the full width at the 1/e intensity and the peak positions.
We define  the spectral delay as the   shift in the
peak position at each energy band compared to the lowest energy 
band (15-25 keV). 

The derived  pulse width and delay are reported in Table 3 and Table 4,
respectively. In Table 3 the
numbers in the brackets are 
the width determined from Norris\textquoteright{} exponential fitting
for individual light curves. Figure 7 shows a comparison between the measured width
following the new method with that determined by fitting the light curve.
We note a systematic decrement of the width measurement by our method
for the first two pulses. This is simply due to the fact that
individual pulses rather than co-added pulses are taken into consideration.
Hence, measured widths are more accurate and devoid of contamination effects from other pulses.

One of the pulse properties of GRB is broadening at lower energies 
(Norris et al. 1996, Hakkila 2011). Figure 8 (left panel) shows the 
pulse width  
\begin{figure}[H]\centering
\includegraphics[angle=0,scale=0.3]{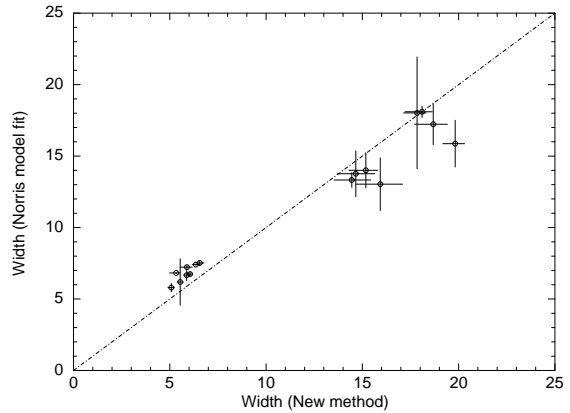}\caption{Comparison
of the width of individual pulses in various energy bands determined
by directly fitting the light curves (Norris model fit) and the three
dimensional model prediction (new method).}

\end{figure}

\begin{table*}\centering
\caption{Variation of the model predicted delay with energy for each pulse. Delay is calculated with respect to
the 15-25 keV band (mean energy: 20.82 keV).}

\begin{tabular}{cccc}
\hline 
Pulse & Energy Channel & Mean Energy (keV) & Delay (s)\tabularnewline
\hline
\hline 
1 & 15-25 vs. 25-50 keV & 35.45 & $-0.135_{-0.022}^{+0.011}$\tabularnewline
 & 15-25 vs. 50-100 keV & 68.07 & $-0.375_{-0.039}^{+0.023}$\tabularnewline
 & 15-25 vs. 100-200 keV & 123.73 & $-0.730_{-0.045}^{+0.043}$\tabularnewline
\hline 
2 & 15-25 vs. 25-50 keV & 35.45 & $-0.195_{-0.005}^{+0.005}$\tabularnewline
 & 15-25 vs. 50-100 keV & 68.07 & $-0.555_{-0.033}^{+0.033}$\tabularnewline
 & 15-25 vs. 100-200 keV & 123.73 & $-1.070_{-0.037}^{+0.023}$\tabularnewline
\hline 
3 & 15-25 vs. 25-50 keV & 35.45 & $-0.015_{0.040}^{+0.041}$\tabularnewline
 & 15-25 vs. 50-100 keV & 68.07 & $-0.040_{-0.030}^{+0.012}$\tabularnewline
 & 15-25 vs. 100-200 keV & 123.73 & $-0.085_{-0.005}^{+0.005}$\tabularnewline
\hline 
4 & 15-25 vs. 25-50 keV & 35.45 & $-0.020_{-0.010}^{+0.010}$\tabularnewline
 & 15-25 vs. 50-100 keV & 68.07 & $-0.065_{-0.040}^{+0.035}$\tabularnewline
 & 15-25 vs. 100-200 keV & 123.73 & $-0.095_{-0.005}^{+0.005}$\tabularnewline
\hline
\end{tabular}
\end{table*} 
variation with energy. While the first two pulses follow
the normal trend, we notice that the width for the third and fourth pulses broadens with
higher energy. This cannot arise from the contamination of the previous pulse
as could have been the argument for fitting the entire light curve.
In our case, each pulse is generated and analyzed separately for the
width determination. The energy-width plot of individual pulses
is fitted with a linear model. The slopes are negative for the first
two pulses ($(-9.4\pm1.7)\times10^{-3}$ s $\rm keV^{-1}$ 
and $(-11.8\pm2.4)\times10^{-3}$ s $\rm keV^{-1}$
respectively; $\chi^2$ = 0.13 (2) and 0.11 (2) respectively)
and positive for the other two ($(18.8\pm11.1)\times10^{-3}$ s $\rm keV^{-1}$
and $(14.2\pm10.6)\times10^{-3}$ s $\rm keV^{-1}$ respectively; $\chi^2$ = 
0.0046 (2) and 0.0021 (2) respectively). Figure 8 (right panel)
shows a comparison of the slopes meaured by the linear fit of data
determined by the two methods (compare slopes determined in section 3.2). We note again that the slopes of
the first two pulses are convincingly negative (low error bar), but
slope of the other two pulses have large error bars. Hence, the evidence
of reverse pulse broadening is tentative, though individual pulse
analysis in the new method shows a better evidence for the reverse
pulse broadening effect (see Figure 8 - left panel).

This deviation from the canonical pulse width broadening with energy decrement may
have a physical significance. In our analysis physical process of pulse emission
is not considered. But phenomenologically, this occurs depending on
the values of $\tau_1$ and $\tau_2$. A detailed
comparison between the first and the third pulse reveals the fact
that they have different pulse width variation albeit having nearly
similar parameters except for  $\tau_{1}$ and $\tau_{2}$. These values
are as follows (the numbers within parentheses are those of the third
pulse): $\alpha$= -1.11 (-1.15), $\beta$ = -2.5 (-2.5), $E_{peak,0}$=359
(324), $\phi_{0}$=12.2 (18.8), $\tau_{1}$=795.4 (353.4), $\tau_{2}$
=0.54 (2.47). It is apparent that the main contributing factors for
the different width-energy relation are the global values of $\tau_{1}$
and $\tau_{2}$. Hence we examined for a generic pulse keeping all
other parameters fixed at the values of the first pulse and varying
$\tau_{1}$ for different values of $\tau_{2}$. 

\begin{figure*}\centering

\includegraphics[angle=0,scale=0.3]{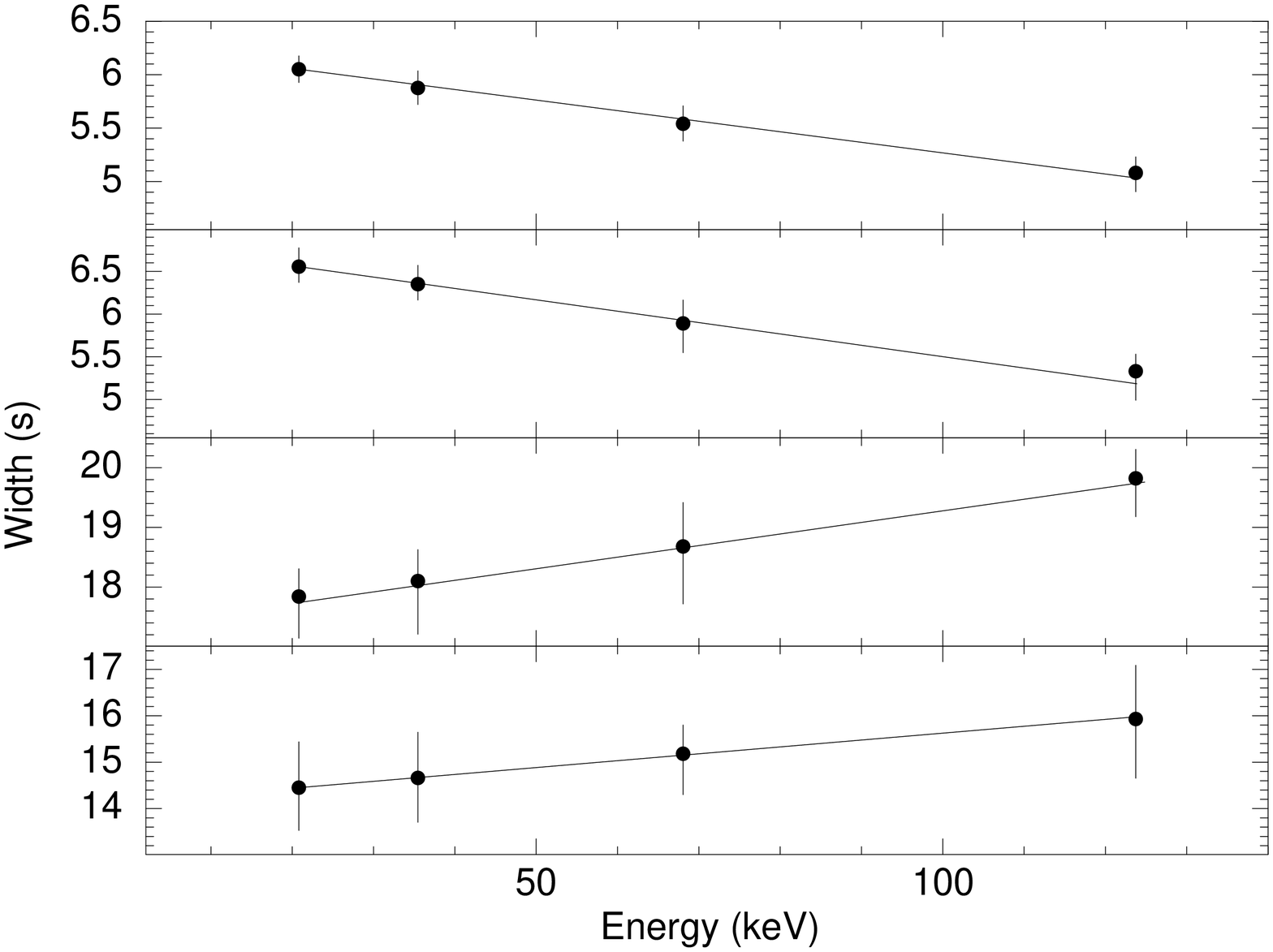}
\includegraphics[angle=0,scale=0.3]{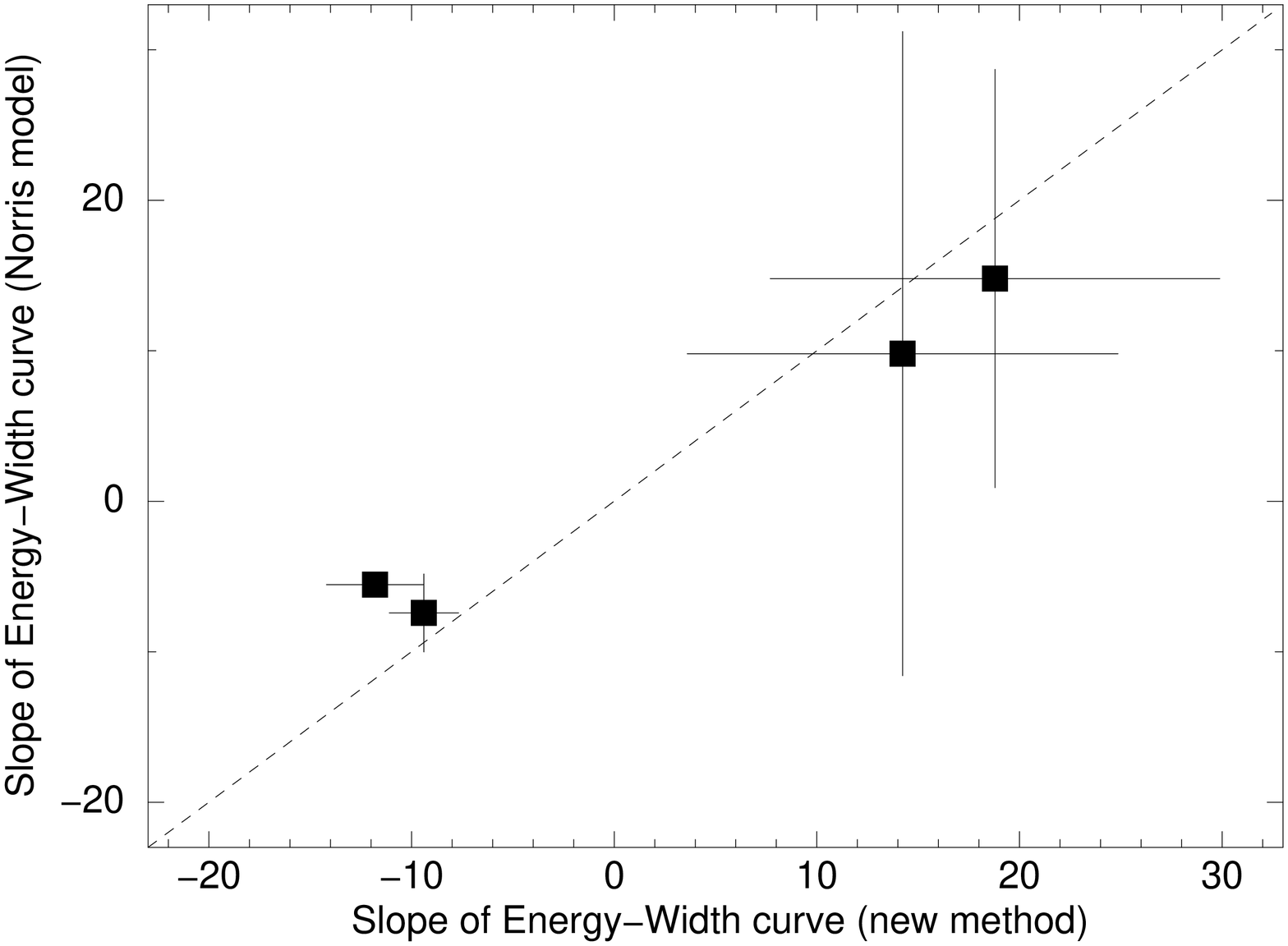}

\caption{(Left panel) Pulse width (w) variation with energy (E) as predicted by our model. 
Pulse 1 to 4 are shown from the top to bottom panels. Width
broadens with lower energy for the first two pulses (normal width broadening) while the
reverse effect is apparent for the third and the fourth pulses (anomalous width broadening).
(Right panel) Comparison of slopes of E-w curves determined by Norris model and the new method.}

\end{figure*}

Figure 9 shows the regions in which pulse broadening or its reverse
phenomenon can occur. This figure is generated by fixing all the parameters
other than $\tau_{1}$ and $\tau_{2}$ to those for the first pulse. For four 
values of $\tau_{2}$ we vary $\tau_{1}$ and note the ratio of the width
in 100-200 keV region ($w_4$) to that in the 15-25 keV ($w_1$) region. The 
solid line parallel to $\tau_{1}$ with $w_{4}$/$w_{1}$=1
divides the plot into two regions: (i) \emph{normal width broadening} region
for which $w_{4}$ < $w_{1}$ and (ii) \emph{anomalous width broadening} region for which
$w_{4}$ > $w_{1}$. Global values of $\tau_{1}$ and $\tau_{2}$
falling in the $w_{4}$ < $w_{1}$ region will lead to normal pulse
broadening (i.e., pulse broadens with lower energy) and vice versa.

We identify from the figure that the $\tau_{1}$ and $\tau_{2}$ values (795.4
and 0.54 respectively) of the first pulse lie well within the region of 
\emph{normal width broadening}. We also note that the values of $\tau_{1}$ and $\tau_{2}$ 
for the third pulse (353.4 and 2.47 respectively) lie within the region of
\emph{anomalous width broadening} despite the fact that the parameter values used,
other than $\tau_{1}$ and $\tau_{2}$, are those of the first pulse. This shows how insensitive
is the pulse broadening effect with all other parameters and crucially
determined by the global values of $\tau_{1}$ and $\tau_{2}$.

Another pulse property is the hard-to-soft spectral evolution. In Figure 10 
we have plotted the model predicted delay as a function of energy (left panel).
%and also have given a comparison of these delays with the observed delays
%(right panel).
It can be seen that the spectral time delay always shows 
a decrement with energy. The hard X ray always precedes
the softer one. The values for the first three peaks can be compared with the
observed delay given in  Rao et al. (2011) and this is shown in the right panel of
Figure 10. The time intervals taken by Rao et al. (2011)
are $T_{0}$ to $T_{0}$+50 s, $T_{0}$+50 s to $T_{0}$+77 s, $T_{0}$+77 s
to $T_{0}$+100 s, $T_{0}$+100 s to $T_{0}$+180 s. Among these the first
time bin contains the precursor burst which is not taken in our calculation.
The first and the second pulses are combined in the second time bin.
The third pulse is covered in the third time bin. Data of the fourth
pulse cannot be compared with the time bin as they differ by $\sim$
55 s (80 s in time bin as compared to $\sim$ 25 s in pulse analysis). 
The deviation of the model predicted delay from the observed delay
can be explained by the fact that the delay in the second time
bin ($T_{0}$+50 to $T_{0}$+77) is the combined effect of the first two
pulses. The data for the third time bin agrees within error 
with the model predicted delay. Also, the observed delays are calculated
using peak position measurement and cross correlation of the pulses 
whereas model predicted delays are calculated from peak position measurement alone.

\subsection{Correlations and the Bigger Picture}

One of the important properties of the prompt emission is that the various measured 
quantities 
\begin{figure}[H]\centering
\includegraphics[angle=0,scale=0.3]{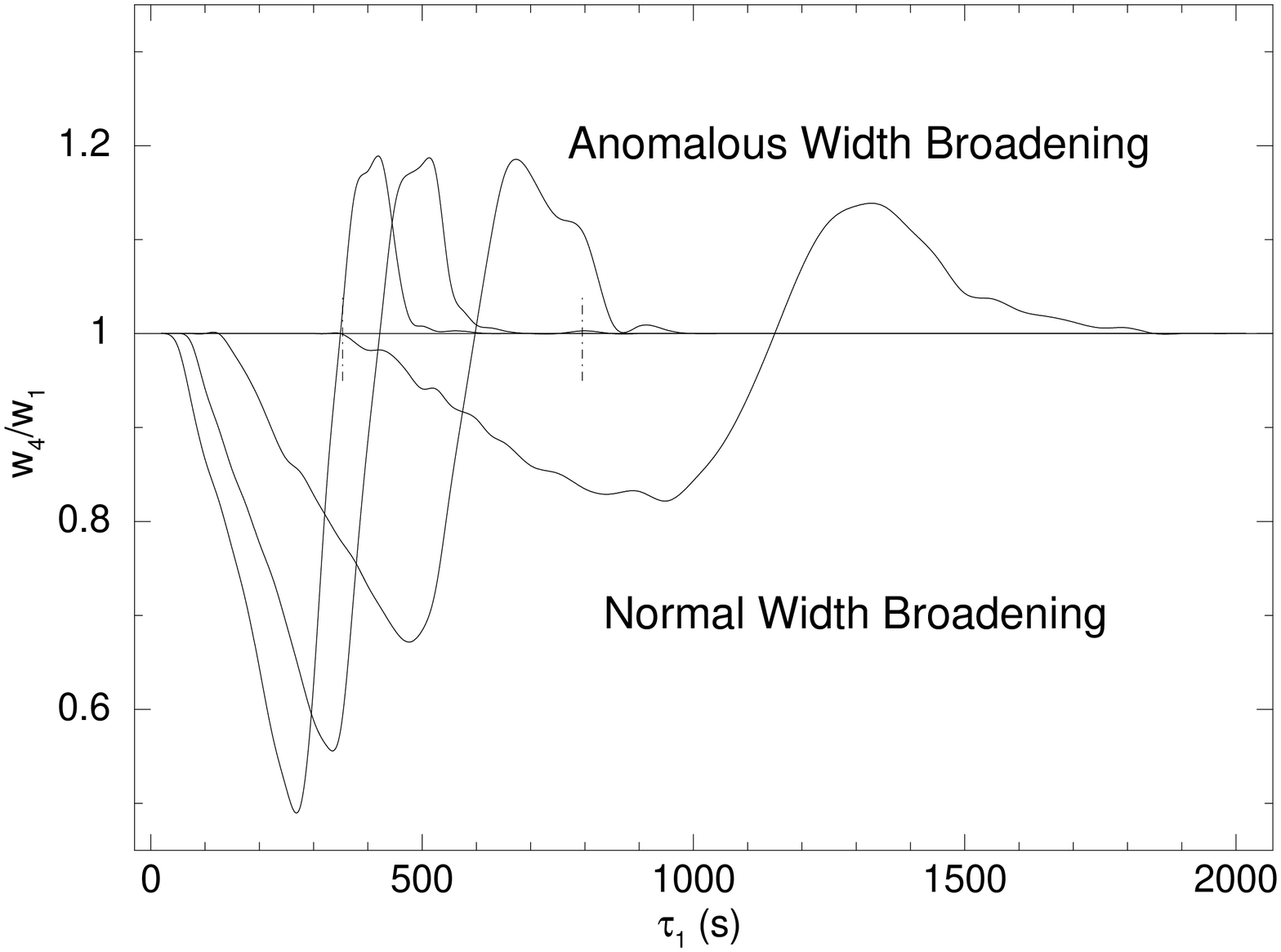}\caption{Ratio of the width in 100-200 keV ($w_{4}$)
to the  width in 15-25 keV ($w_{1}$), as predicted by the pulse model, 
is plotted against $\tau_{1}$ for different values of $\tau_{2}$. In the figure
maximum as well as the minimum of the curves move to lower $\tau_{1}$ values for higher $\tau_{2}$.
The values of $\tau_{2}$ are 0.54, 1.20, 1.90, and 2.47, respectively from left to right.
The $\tau_{1}$ values determined for the second and the third pulses of
the GRB  (795.4 s and 353.4 s, respectively) are shown by dash-dotted lines.
%The width-energy relation is crucially determined by the initial values
%of $\tau_{1}$ and $\tau_{2}$. 
The solid line 
%parallel to the $\tau_{1}$ axis with y-intercept equal to 1 
divides the region into two parts: the Normal width broadening region
and the Anomalous width broadening region (see section 4.2.2).}

\end{figure}

\begin{figure*}\centering
\subfloat{\includegraphics[angle=0,scale=0.3]{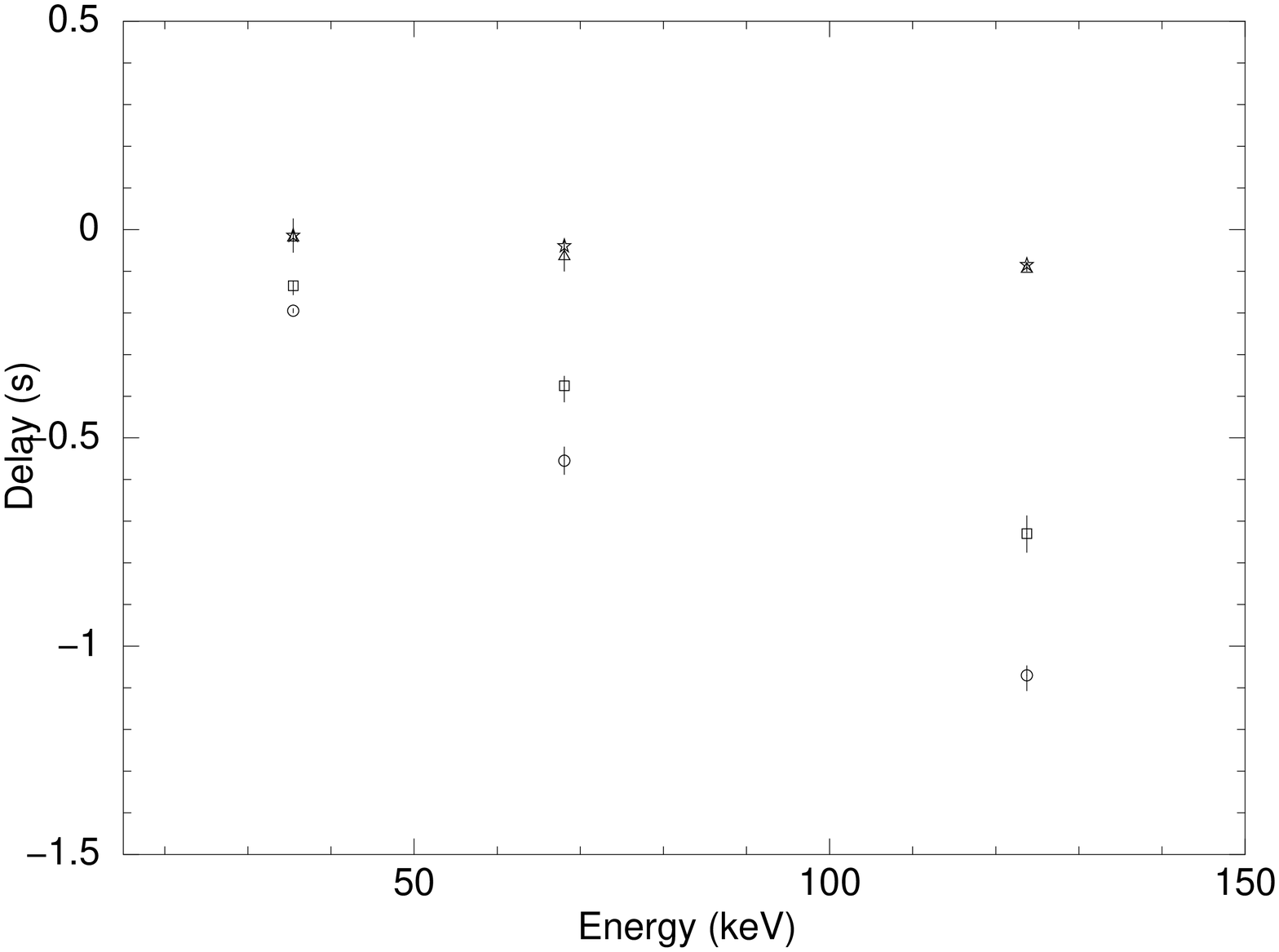}}\subfloat{\includegraphics[angle=0,scale=0.3]{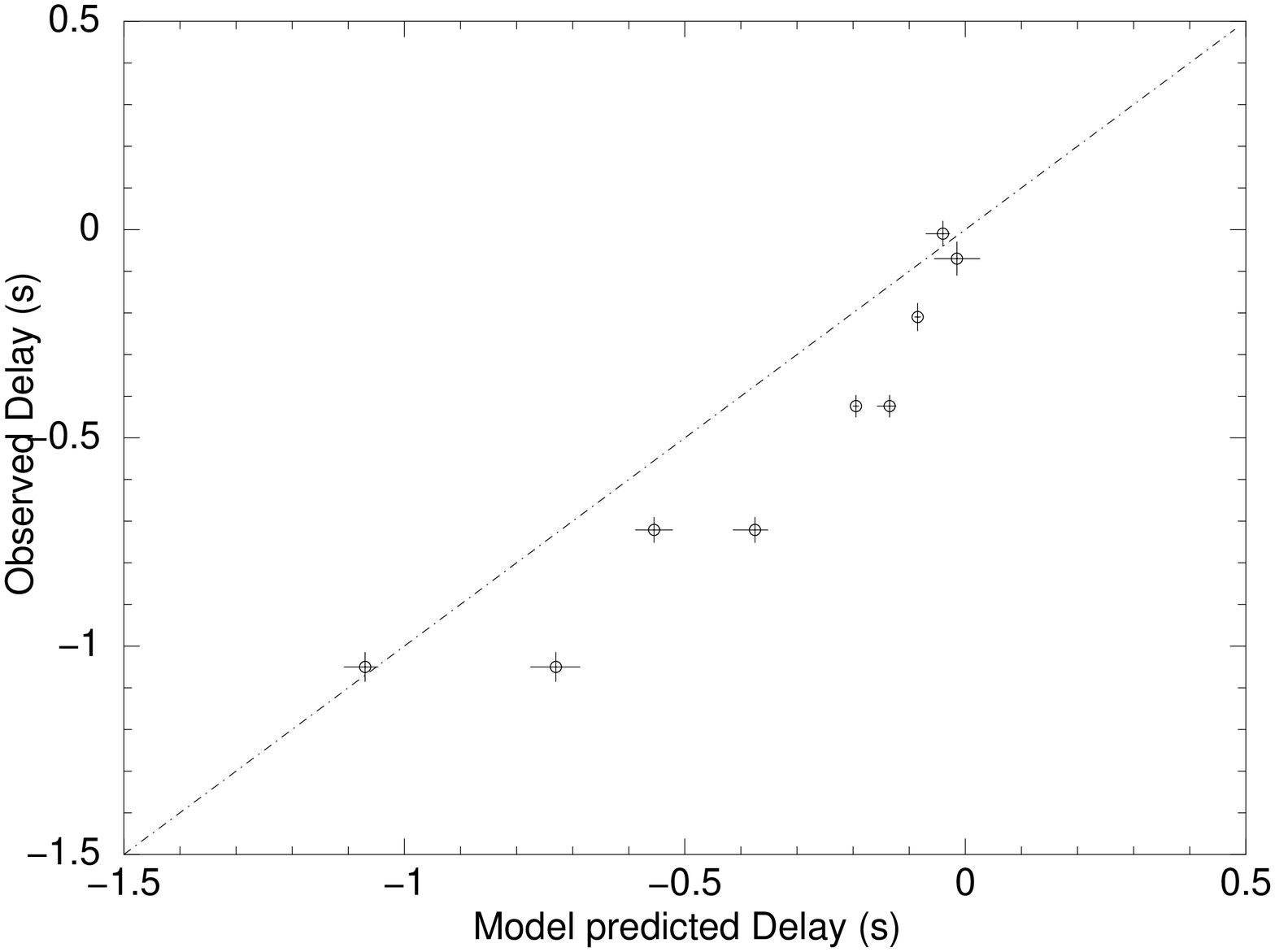}

}\caption{Model predicted delay of light curve with mean energy
E, with respect to the 15-25 keV light curve (left panel). For the four pulses
the data are shown in square, circle, triangle, and star respectively.
Right panel shows the comparison of the measured delay (Rao et
al. 2011) with the model predicted delay. The lower six points
are model prediction for first and second pulses compared with the
50-77 sec interval (see text). The end points along the
x-axis for this data group are the model predicted delays with
the left one for the second pulse. The measured delay along with error is shown
in the y axis. The straight line is the line of equality. }

\end{figure*}
correlate with luminosity. Particularly, the peak energy
of the $\nu F_{\nu}$ spectrum, $E_{peak}$ correlates with isotropic
energy $E_{iso}$ (Amati et al. 2002) and also with the  peak isotropic luminosity $L_{iso}$
(Yonetoku et al. 2004). Ghirlanda et al. (2004) showed that the correlation
is tighter with collimation-corrected energy, $E_{\gamma}$. Schaefer
(2007) used a set of 69 GRB sample (up till 7 June 2006) with measured
redshift to calculate various correlations viz. lag-luminosity ($\tau_{lag}-L$),
variability-luminosity ($V-L$), $E_{peak}-L$, $E_{peak}-E_{\gamma}$,
 minimum rise time-luminosity ($\tau_{RT}-L$), number of peaks-luminosity
($N_{peak}-L$) to get distance moduli for each. Weighted average
of them gave the distance modulus ($\mu$) which is plotted against
the measured redshift (z) extending the Hubble Diagram (HD) to z >6.
Therefore, these correlations can be used to standardize the GRB energy
budget making GRBs a class of cosmic distance indicators. Also, these
correlations might point to a underlying fundamental physical process 
(Yamazaki et al. 2004; Rees \& M{\'e}sz{\'a}ros 2005; Xu et al. 2005; 
Firmani et al. 2005; 2006; 2007; Ghirlanda et al. 2006a;b; 
Wang et al. 2006; Thompson 2006; Thompson et al. 2007; 
Liang et al. 2005; 2008 and the references therein; Li et al. 2008; Qi et al. 2008).

In order to establish the reality of these correlations one must show
that all of these correlations hold just as the same way within a
single GRB as for a sample of many GRBs. Ghirlanda et al. (2010) showed
for a set of GRBs that the \emph{time-resolved} correlations of a
single GRB are similar to the \emph{time integrated} correlations
among different GRBs. In the present work we have used the underlying
pulse structure to investigate the correlations: $E_{peak,0}-L_{iso}$,
$E_{peak,0}-E_{\gamma,iso}$, $\phi_{0}-L_{iso}$ and $\phi_{0}-E_{\gamma,iso}$
within the pulses of a single GRB.

\begin{figure*}\centering
\subfloat{\includegraphics[angle=0,scale=0.3]{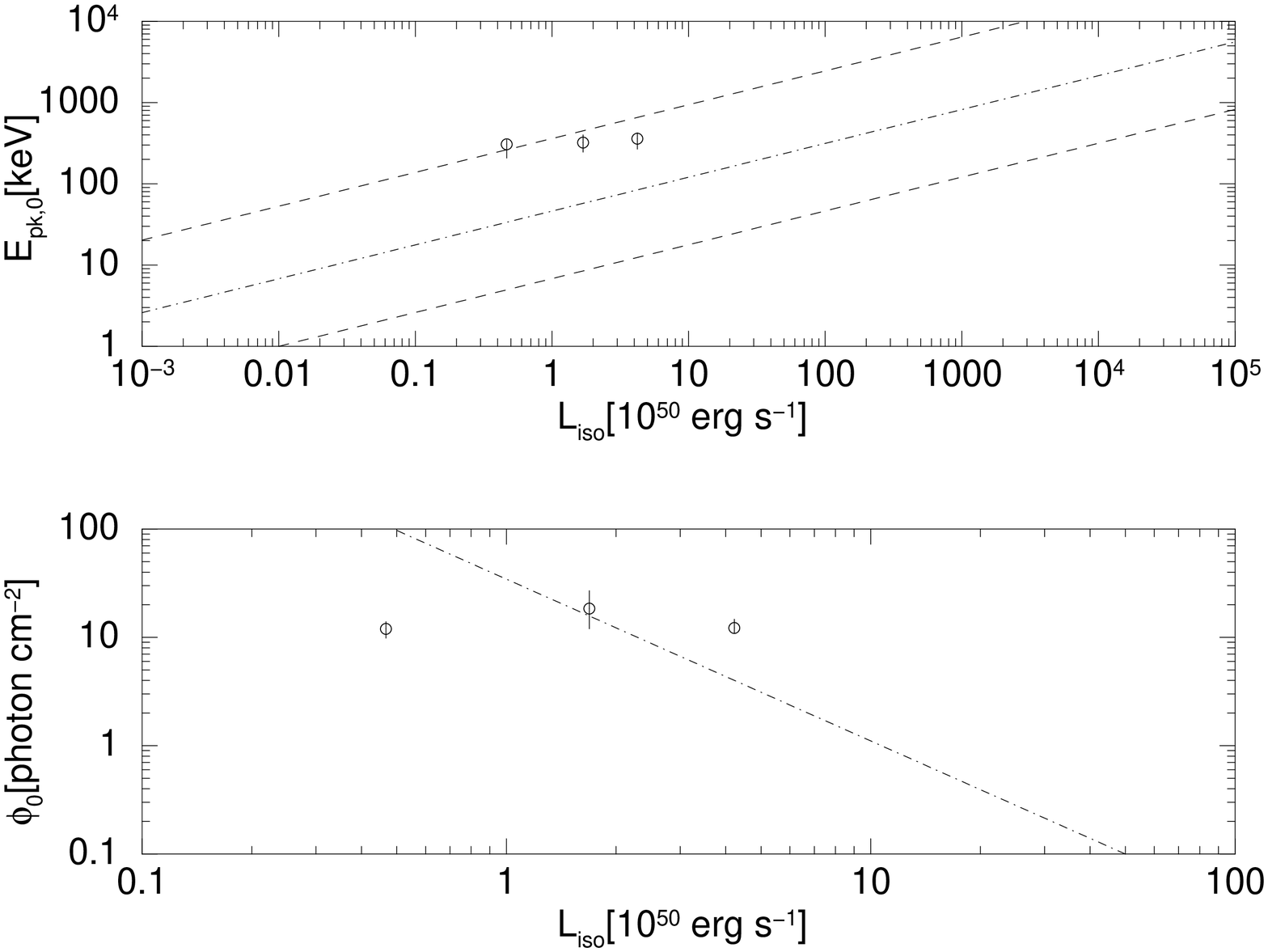}}\subfloat{\includegraphics[angle=0,scale=0.3]{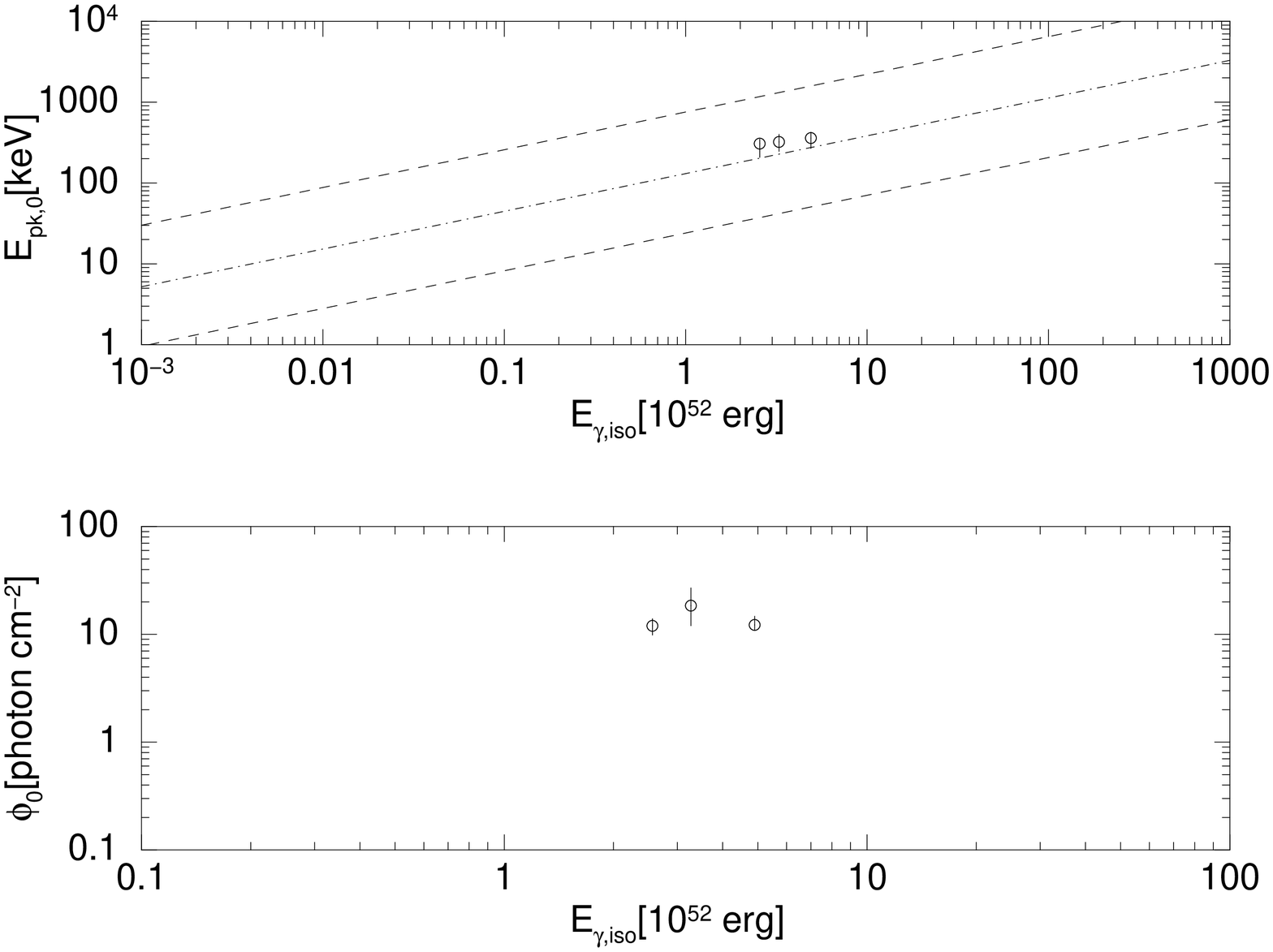}

}\caption{Correlations between (a)$E_{peak,0}-L_{iso}$ (left panel, top); the dot-dashed
line is the mean $E_{peak}-L_{iso}$ curve while the dashed lines
represent 3$\sigma$ scatter (from Ghirlanda et al. 2010), $\phi_{0}-L_{iso}$
(bottom); the dot-dashed line is the mean $\phi_{0}-L_{iso}$ line
(from Kocevski et al. 2003) and (b) $E_{peak,0}-E_{\gamma,iso}$ (right
panel, top); the dot-dashed line is the mean $E_{peak}-E_{\gamma,iso}$
curve while the dashed lines represent 3$\sigma$ scatter (from Ghirlanda
et al. 2010), $\phi_{0}-E_{\gamma,iso}$ (bottom).}

\end{figure*}

The correlations are shown in Figure 11. The mean and the 3$\sigma$ scatter 
of $E_{peak}-L_{iso}$ (top left panel) and $E_{peak}-E_{\gamma,iso}$ 
(top right panel; Ghirlanda et al. 2010) are shown for reference. 
As expected, the $E_{peak,0}$ values always lie above 
the mean of $E_{peak}$ values. It seems that
the correlations of $E_{peak,0}$ with $L_{iso}$ and $E_{\gamma,iso}$ are rather tighter in
our case, though with only three points we cannot draw any definite conclusions. 
While the parameter $E_{peak,0}$ is a constant, $E_{peak}$
is an average quantity. Hence, we expect that $E_{peak,0}$ provides 
a better pulse description than $E_{peak}$. This can be tested in future for correlations of $E_{peak,0}$
rather than $E_{peak}$ with other quantities in different GRBs. 
%The
%data set is too small (only three) to infer the correlations right away. 
But if the
same correlations hold for different GRBs then we might conclude (a)
the correlations are rather physical and not an outcome of the bias
in sample selection, (b) in a single multi-peaked GRB each pulse is
generated independently. We also note that the calculated $L_{iso}$ and $E_{\gamma,iso}$
are somewhat lower than those calculated by Ghirlanda et al. (2010).
This must have come from the instrumental effect as noticed earlier.
The normalization of BAT is always less than that of \emph{Fermi}. This leads
to a lower flux calculated from the BAT spectrum.

\section{DISCUSSION AND CONCLUSIONS}

We have attempted to explain the timing and spectral variations within
a pulse of GRB by a simple empirical model and have given a prescription
to measure the parameters of this model. We use simultaneous data
from \emph{Swift}/BAT and \emph{Fermi}/GBM to determine the average pulse spectral
characteristics using the Band model. Then we use the integrated light
curve from \emph{Swift}/BAT to get a simple description of the time evolution
using Norris\textquoteright{} exponential model. Using these parameters,
we develop a method to fit the spectrum using XSPEC table model by
assuming that $E_{peak}(t)$ decreases exponentially with fluence  in the
pulse. We determine the parameters of the model and then find that
these parameters can correctly describe the energy dependence of the
pulse characteristics like width and spectral lags. We also find, in the data,
a tentative evidence of \emph{anomalous width broadening with energy}
for some pulses. The joint spectral and timing analysis developed here
confirms this phenomena for the same pulses at least in the lower energy bands.

The method developed here is applicable for both single pulse
 and structured GRBs. Most of the long GRBs are structured which makes them
 difficult for pulse analysis. As far as the correlations are concerned,
 pulse analysis is more indicative than intensity guided time-resolved 
spectroscopy. Hence to draw conclusion regarding these correlations 
within a sample of GRBs and within the pulses of a single GRB we must include 
these structured GRBs with multiple pulses. Kocevski et al. (2003) have prescribed 
a method for short, clean pulses. Our method extends the domain to all kinds of GRBs. 
Also the present method is much more  physical than the time-resolved one in the sense that 
the individual pulses are treated separately. Hence there won't be 
contamination effects. The L$_{iso}$-E$_{peak,0}$ looks tighter because correlations 
within pulses may be more physical than the time-resolved correlation 
or the time integrated correlations. 
In future the  method developed here can be used for a large set 
of GRBs (onto their individual pulses) to investigate the improvement in the correlations.

It is very interesting to note that the two distinct pulses (pulse 3 and pulse 4)
in GRB 090618 have E$_{iso}$ values within a factor of 2, but widely different
E$_{peak}$ values (114 keV and 33 keV, respectively). They will show a huge 
scatter if used individually for a global correlation. The E$_{peak,0}$ values,
on the other hand, are closer to each other and lead to a lower 
scatter in the correlation (see Figure 11, top right panel). Hence we suggest that
the empirical relations are intrinsic to the emission mechanism of GRBs and 
E$_{peak,0}$ values, instead of E$_{peak}$, should be the correct parameter to be
used for correlation studies.

The method described in this paper is quite generic in the sense that any
empirical description of the spectrum and its evolution with time can be tested
against the data and the derived parameters can be used for global  correlation studies.
For example the Band spectrum is derived from an extensive study
of data from \emph{CGRO}/BATSE, which is primarily sensitive above 
50 keV (Band et al. 1993). Extension of the data to lower energies
in fact requires additional features. For example a time resolved 
spectral study of GRB~041006 using \emph{HETE-2} data covering the energy
range of 2 keV -- 400 keV, indicated the presence of soft components,
each of them having distinctive time evolution (Shirasaki et al. 2008). 
We emphasize again here that the
use of different instruments can help in pinning down the systematics, 
identifying additional spectral components and developing an empirical 
spectral formalism. The simultaneous data from
\emph{Swift}/BAT and \emph{Fermi}/GBM have  the potential to explore the
existence of additional components because the highly sensitive BAT has an 
energy range ($\sim$15 keV -- 200 keV) which is a subset of the GBM
energy range.  Our joint analysis of BAT and GBM data
has shown that in the 20 -- 40 keV region there is a distinct discrepancy.
This could be due to a) systematic errors in each of the detectors, b)
presence of additional components or c) an interplay of both these.
Since the spectral fitting method used in  XSPEC is forward in nature 
(that is fold the assumed spectrum and compare with the data - see Arnaud 1996),
the derived spectral parameters will have strong bias of the bandwidth of the
detectors, particularly for such low resolution spectral data.  
%the be strongly biased always strhas the habit of \emph{following} the 
%input model for such low resolution spectral data, 
Hence a joint fitting of 
data with good overlap will result in a better estimate of the parameters.
% value with a minimum of uncertainties.

In conclusion, we have developed a method to test empirical descriptions of
the energy spectra and their evolution with time in GRBs and to derive the
underlying physical parameters. This is
the first attempt in this different approach to study various parameters
and predict quantities like spectral delay and pulse width of individual pulses
in a GRB.

\section{Acknowledgments} This research has made use of data obtained through the
HEASARC Online Service, provided by the NASA/GSFC, in support of NASA High Energy
Astrophysics Programs. We thank the referee for the valuable suggestions 
in making the text more readable and quantifying some of the results.


\begin{thebibliography}{}

\bibitem[Amati et al.(2002)]{2002A&A...390...81A} Amati, L., et al.\ 2002, \aap, 390, 81 
\bibitem[Band et al.(1993)]{1993ApJ...413..281B} Band, D., et al.\ 1993, \apj, 413, 281 
\bibitem[Barthelmy et al.(2005)]{2005SSRv..120..143B} Barthelmy, S.~D., et al.\ 2005, \ssr, 120, 143 
\bibitem[Beardmore \& Schady(2009)]{2009GCN..9528....1B} Beardmore, A.~P., \& Schady, P.\ 2009, GRB Coordinates Network, 9528, 1 
\bibitem[Cenko(2009)]{2009GCN..9513....1C} Cenko, S.~B.\ 2009, GRB Coordinates Network, 9513, 1 
\bibitem[Cenko et al.(2009)]{2009GCN..9518....1C} Cenko, S.~B., Perley, D.~A., Junkkarinen, V., Burbidge, M., Diego, U.~S., \& Miller, K.\ 2009, GRB Coordinates Network, 9518, 1 
\bibitem[Fenimore \& Ramirez-Ruiz(2000)]{2000astro.ph..4176F} Fenimore, E.~E., \& Ramirez-Ruiz, E.\ 2000, arXiv:astro-ph/0004176 
\bibitem[Firmani et al.(2007)]{2007RMxAA..43..203F} Firmani, C., Avila-Reese, V., Ghisellini, G., \& Ghirlanda, G.\ 2007, \rmxaa, 43, 203 
\bibitem[Firmani et al.(2006)]{2006MNRAS.372L..28F} Firmani, C., Avila-Reese, V., Ghisellini, G., \& Ghirlanda, G.\ 2006, \mnras, 372, L28 
\bibitem[Firmani et al.(2005)]{2005MNRAS.360L...1F} Firmani, C., Ghisellini, G., Ghirlanda, G., \& Avila-Reese, V.\ 2005, \mnras, 360, L1 
\bibitem[Fryer et al.(1999)]{1999ApJ...526..152F} Fryer, C.~L., Woosley, S.~E., \& Hartmann, D.~H.\ 1999, \apj, 526, 152 
\bibitem[Gehrels et al.(2004)]{2004ApJ...611.1005G} Gehrels, N., et al.\ 2004, \apj, 611, 1005 
\bibitem[Ghirlanda et al.(2006)]{2006NJPh....8..123G} Ghirlanda, G., Ghisellini, G., \& Firmani, C.\ 2006, New Journal of Physics, 8, 123 
\bibitem[Ghirlanda et al.(2006)]{2006A&A...452..839G} Ghirlanda, G., Ghisellini, G., Firmani, C., Nava, L., Tavecchio, F., \& Lazzati, D.\ 2006, \aap, 452, 839 
\bibitem[Ghirlanda et al.(2010)]{2010A&A...511A..43G} Ghirlanda, G., Nava, L., \& Ghisellini, G.\ 2010, \aap, 511, A43 
\bibitem[Ghirlanda et al.(2004)]{2004ApJ...616..331G} Ghirlanda, G., Ghisellini, G., \& Lazzati, D.\ 2004, \apj, 616, 331 
\bibitem[Golenetskii et al.(2009)]{2009GCN..9553....1G} Golenetskii, S., et al.\ 2009, GRB Coordinates Network, 9553, 1 
\bibitem[Hakkila \& Preece(2011)]{2011arXiv1103.5434H} Hakkila, J., \& Preece, R.~D.\ 2011, arXiv:1103.5434 
\bibitem[Kocevski \& Liang(2003)]{2003ApJ...594..385K} Kocevski, D., \& Liang, E.\ 2003, \apj, 594, 385 
\bibitem[Kocevski et al.(2003)]{2003ApJ...596..389K} Kocevski, D., Ryde, F., \& Liang, E.\ 2003, \apj, 596, 389 
\bibitem[Kono et al.(2009)]{2009GCN..9568....1K} Kono, K., et al.\ 2009, GRB Coordinates Network, 9568, 1 
\bibitem[Kouveliotou et al.(1993)]{1993ApJ...413L.101K} Kouveliotou, C., Meegan, C.~A., Fishman, G.~J., Bhat, N.~P., Briggs, M.~S., Koshut, T.~M., Paciesas, W.~S., \& Pendleton, G.~N.\ 1993, \apjl, 413, L101 
\bibitem[Li et al.(2008)]{2008ApJ...680...92L} Li, H., Xia, J.-Q., Liu, J., Zhao, G.-B., Fan, Z.-H., \& Zhang, X.\ 2008, \apj, 680, 92 
\bibitem[Liang \& Kargatis(1996)]{1996Natur.381...49L} Liang, E., \& Kargatis, V.\ 1996, \nat, 381, 49 
\bibitem[Liang \& Zhang(2005)]{2005ApJ...633..611L} Liang, E., \& Zhang, B.\ 2005, \apj, 633, 611 
\bibitem[Liang et al.(2008)]{2008ApJ...685..354L} Liang, N., Xiao, W.~K., Liu, Y., \& Zhang, S.~N.\ 2008, \apj, 685, 354 
\bibitem[Longo et al.(2009)]{2009GCN..9524....1L} Longo, F., et al.\ 2009, GRB Coordinates Network, 9524, 1 
\bibitem[M{\'e}sz{\'a}ros(2006)]{2006RPPh...69.2259M} M{\'e}sz{\'a}ros, P.\ 2006, Reports on Progress in Physics, 69, 2259 
\bibitem[M{\'e}sz{\'a}ros(2006)]{2006sgrb.progE..15M} M{\'e}sz{\'a}ros, P.\ 2006, KITP Program: The Supernova Gamma-Ray Burst Connection,  
\bibitem[McBreen(2009)]{2009GCN..9535....1M} McBreen, S.\ 2009, GRB Coordinates Network, 9535, 1 
\bibitem[Meegan(2009)]{2009APS..APR.T4001M} Meegan, C.\ 2009, APS Meeting Abstracts, 4001 
\bibitem[Meszaros \& Rees(1997)]{1997ApJ...482L..29M} Meszaros, P., \& Rees, M.~J.\ 1997, \apjl, 482, L29 
\bibitem[Meszaros \& Rees(1992)]{1992ApJ...397..570M} Meszaros, P., \& Rees, M.~J.\ 1992, \apj, 397, 570 
\bibitem[Metzger(2010)]{2010ASPC..432...81M} Metzger, B.~D.\ 2010, New Horizons in Astronomy: Frank N.~Bash Symposium 2009, 432, 81 
\bibitem[Metzger et al.(2007)]{2007AIPC..937..521M} Metzger, B.~D., Thompson, T.~A., \& Quataert, E.\ 2007, Supernova 1987A: 20 Years After: Supernovae and Gamma-Ray Bursters, 937, 521 
\bibitem[Nava et al.(2011)]{2011A&A...530A..21N} Nava, L., Ghirlanda, G., Ghisellini, G., \& Celotti, A.\ 2011, \aap, 530, A21 
\bibitem[Nemiroff(2000)]{2000ApJ...544..805N} Nemiroff, R.~J.\ 2000, \apj, 544, 805 
\bibitem[Norris et al.(2005)]{2005ApJ...627..324N} Norris, J.~P., Bonnell, J.~T., Kazanas, D., Scargle, J.~D., Hakkila, J., \& Giblin, T.~W.\ 2005, \apj, 627, 324 
\bibitem[Norris et al.(1996)]{1996ApJ...459..393N} Norris, J.~P., Nemiroff, R.~J., Bonnell, J.~T., Scargle, J.~D., Kouveliotou, C., Paciesas, W.~S., Meegan, C.~A., \& Fishman, G.~J.\ 1996, \apj, 459, 393 
\bibitem[Norris et al.(2000)]{2000ApJ...534..248N} Norris, J.~P., Marani, G.~F., \& Bonnell, J.~T.\ 2000, \apj, 534, 248 
\bibitem[Paczynski(1986)]{1986ApJ...308L..43P} Paczynski, B.\ 1986, \apjl, 308, L43 
\bibitem[Paczynski(1998)]{1998ApJ...494L..45P} Paczynski, B.\ 1998, \apjl, 494, L45 
\bibitem[Perley(2009)]{2009GCN..9514....1P} Perley, D.~A.\ 2009, GRB Coordinates Network, 9514, 1 
\bibitem[Qi et al.(2008)]{2008A&A...483...49Q} Qi, S., Wang, F.-Y., \& Lu, T.\ 2008, \aap, 483, 49 
\bibitem[Rao et al.(2009)]{2009GCN..9665....1R} Rao, A.~R., et al.\ 2009, GRB Coordinates Network, 9665, 1 
\bibitem[Rao et al.(2011)]{2011ApJ...728...42R} Rao, A.~R., et al.\ 2011, \apj, 728, 42 
\bibitem[Rees \& M{\'e}sz{\'a}ros(2005)]{2005ApJ...628..847R} Rees, M.~J., \& M{\'e}sz{\'a}ros, P.\ 2005, \apj, 628, 847 
\bibitem[Rosswog \& Ramirez-Ruiz(2003)]{2003MNRAS.343L..36R} Rosswog, S., \& Ramirez-Ruiz, E.\ 2003, \mnras, 343, L36 
\bibitem[Rosswog et al.(2003)]{2003MNRAS.345.1077R} Rosswog, S., Ramirez-Ruiz, E., \& Davies, M.~B.\ 2003, \mnras, 345, 1077 
\bibitem[Sakamoto et al.(2011)]{2011PASJ...63..215S} Sakamoto, T., et al.\ 2011, \pasj, 63, 215 
\bibitem[Schady(2009)]{2009GCN..9527....1S} Schady, P.\ 2009, GRB Coordinates Network, 9527, 1 
\bibitem[Schady et al.(2009)]{2009GCNR..232....1S} Schady, P., Baumgartner, W.~H., \& Beardmore, A.~P.\ 2009, GCN Report, 232, 1 
\bibitem[Schady et al.(2009)]{2009GCN..9512....1S} Schady, P., et al.\ 2009, GRB Coordinates Network, 9512, 1 
\bibitem[Schaefer(2002)]{2002cosp...34E1141S} Schaefer, B.\ 2002, 34th COSPAR Scientific Assembly, 34,  
\bibitem[Schaefer(2007)]{2007ApJ...660...16S} Schaefer, B.~E.\ 2007, \apj, 660, 16 
\bibitem[Shirasaki et al.(2008)]{2008PASJ...60..919S} Shirasaki, Y., et al.\ 2008, \pasj, 60, 919 
\bibitem[Thompson et al.(2007)]{2007ApJ...666.1012T} Thompson, C., M{\'e}sz{\'a}ros, P., \& Rees, M.~J.\ 2007, \apj, 666, 1012 
\bibitem[Thompson(2006)]{2006ApJ...651..333T} Thompson, C.\ 2006, \apj, 651, 333 
\bibitem[van Paradijs et al.(2000)]{2000ARA&A..38..379V} van Paradijs, J., Kouveliotou, C., \& Wijers, R.~A.~M.~J.\ 2000, \araa, 38, 379 
\bibitem[Wang \& Dai(2006)]{2006MNRAS.368..371W} Wang, F.~Y., \& Dai, Z.~G.\ 2006, \mnras, 368, 371 
\bibitem[Woosley(1993)]{1993ApJ...405..273W} Woosley, S.~E.\ 1993, \apj, 405, 273 
\bibitem[Xu et al.(2005)]{2005ApJ...633..603X} Xu, D., Dai, Z.~G., \& Liang, E.~W.\ 2005, \apj, 633, 603 
\bibitem[Yamazaki et al.(2004)]{2004ApJ...606L..33Y} Yamazaki, R., Ioka, K., \& Nakamura, T.\ 2004, \apjl, 606, L33 
\bibitem[Yonetoku et al.(2004)]{2004ApJ...609..935Y} Yonetoku, D., Murakami, T., Nakamura, T., Yamazaki, R., Inoue, A.~K., \& Ioka, K.\ 2004, \apj, 609, 935 

\end{thebibliography}
\end{document}